\documentstyle[11pt,aaspp4]{article}
\tighten
\begin{document}
\title{The Active Corona of HD~35850 (F8~V)}
\author{Marc Gagn\'e, Jeff A. Valenti, and Jeffrey L. Linsky}
\affil{JILA, University of Colorado and NIST, Boulder, CO 80309-0440}

\author{Gianpiero Tagliaferri and Stefano Covino}
\affil{Osservatorio Astronomico di Brera, via E. Bianchi 46, I-22055 Merate,Italia}

\and

\author{Manuel G\"udel}
\affil{Paul Scherrer Institute, CH-5232, Villigen PSI, Switzerland} 

\received{ 1998 June 22 } \accepted{ 1998 November 4 }

\slugcomment{To appear in The Astrophysical Journal}

\begin{abstract}

We present {\it Extreme Ultraviolet Explorer} spectroscopy and
photometry of the nearby F8~V star HD~35850 (HR~1817).  The {\it EUVE}
short-wavelength 75--175~\AA\ and medium-wavelength 160--365~\AA\
spectra reveal 28 emission lines from \ion{Fe}{9} and 
\ion{Fe}{15} to \ion{Fe}{24}.
The \ion{Fe}{21} $\lambda\lambda 102, 129$ ratio yields an upper limit for
the coronal electron density, $\log n_{\rm e} < 11.6$~cm$^{-3}$.
The {\it EUVE} SW spectrum shows a small but clearly detectable continuum.
The 75--150~\AA\ line-to-continuum ratio indicates approximately solar Fe
abundances, with $0.8 < Z < 1.6$ (90\% confidence interval).  
Upper limits have been derived for a dozen high-emissivity
\ion{Fe}{10} through \ion{Fe}{14} lines.
The resulting EM distribution is characterized
by two broad temperature components at $\log T$ of 6.8 and 7.4. 
Over the course of the 1-week observation, large-amplitude, long-duration
flares were not seen in the {\it EUVE} Deep Survey light curve,
though the light curve does show signs of persistent, low-level flaring,
and possible rotational modulation. 

The {\it EUVE} spectra have been compared with non-simultaneous {\it ASCA}
SIS spectra of HD~35850 obtained in 1995.  The SPEX DEM analysis of
the SIS spectrum indicate the same temperature distribution as the {\it EUVE}
DEM analysis.  However, the SIS spectra suggest sub-solar abundances,
$0.34 < Z < 0.81$.
Although some of the discrepancy may be the result of incomplete X-ray line
lists, we cannot explain the disagreement between the {\it EUVE}
line-to-continuum ratio and the {\it ASCA}-derived Fe abundance.

The X-ray surface flux on HD~35850 is 
comparable to that of cooler dwarfs of comparable age and rotation like
EK~Draconis (G0~V) and AB~Doradus (K1~V). Given its youth ($t \approx 100$~Myr),
its rapid rotation ($v \sin i \approx  50$~km~s$^{-1}$), and its high X-ray 
activity ($L_{\rm X} \approx 1.5 \times 10^{30}$~ergs~s$^{-1}$),
HD~35850 may represent an activity extremum for single,
main-sequence F-type stars.  The variability and EM distribution 
can be reconstructed using the continuous flaring model of G\"udel
provided that the flare distribution has a power-law index
$\alpha \approx 1.8$. Similar results obtained for other young solar
analogs suggest that continuous flaring is a viable
coronal heating mechanism on rapidly rotating, late-type,
main-sequence stars. 

\end{abstract}


\section{Introduction}

By analogy to the Sun, magnetic heating in the chromospheres,
transition regions, and coronae of late-type stars is observed as
cooling in the Balmer lines, the \ion{Ca}{2} lines, and in UV, FUV,
EUV, and soft X-ray continuum and line emission.  Results from the
Mt. Wilson \ion{Ca}{2}~H~and~K survey of single lower main-sequence
field stars has provided compelling evidence of Solar-like
chromospheric active regions, activity cycles, differential rotation,
and occasional Maunder minima in mid-F to early-M stars (e.g., Noyes
et al. 1984; Baliunas et al. 1996; Gray \& Baliunas 1997; Baliunas et
al. 1995).

Coronal heating and magnetic dynamos on the Sun and in stars have been
discussed extensively in the literature (cf. Haisch \& Schmitt 1996).
The various flares,
microflares, nanoflares, and brightenings observed on the Sun are
thought to be manifestations of current sheet reconnection and much
of the coronal heating debate is focussed on whether or not these
transient events can account for the radiative output of the solar corona
(e.g., Oreshina \& Somov 1998). 
While both continuous (e.g., MHD waves) and stochastic (nanoflaring)
processes probably contribute to magnetic coronal heating, the
dominant process has yet to be identified.  

Empirically, coronal emission on active, cool, main-sequence
stars is similar to flare emission from solar active regions.
For example, Benz \& G\"udel (1994) find a similar correlation between
X-ray luminosity and non-thermal continuum radio luminosity for solar
flares and active stars.  On the Sun, mildly relativistic electrons 
accelerated along coronal magnetic field lines produce gyrosynchrotron
radio emission. The presence of persistent but variable nonthermal
radio emission from cool stars indicates a continual replenishment of the
relativistic electron population while a correlation between coronal
and radio emissions suggests a causal relationship between electron
acceleration in magnetically confined loops and coronal heating.
In the 50,000--200,000~K regime, further evidence of continual flaring
comes from the broad emission-line components seen in high-resolution UV
spectra of active dwarfs and RS~CVn binaries.
Because the broad-component line widths are 2--4 times the
thermal width and because the broad components are blue-shifted
relative to the narrow components, Wood et al. (1996) suggest that the
broad components are produced by transition-region 
explosive events.

Two well-studied examples of extreme main-sequence magnetic activity are the
Pleiades-age, rapid rotators AB~Doradus (K0~V) and EK~Draconis (G0~V).
Spectro-polarimetric monitoring of AB~Dor shows long-lived, cool magnetic spots
and latitudinal differential rotation (Donati \& Collier Cameron 1997).
{\it ROSAT} PSPC soft X-ray photometry of AB~Dor (K\"urster et
al. 1997) shows rotationally modulated flare and quiescent emission
with the same phase and period (12.4~h) as the photospheric spots.
{\it ASCA} and {\it EUVE} spectra of AB~Dor and EK~Dra
indicate coronal emission-measure distributions, ${\rm Em}(T)$,
with peaks at 5--8~MK and 20--30~MK
(Mewe et al. 1996; G\"udel et al. 1997).

The 5--8~MK component has been observed in other moderately active
coronal sources, most notably Capella (Brickhouse, Raymond, \& Smith 1995).
As has been pointed out by Gehrels \& Williams (1993), optically thin
plasma tends to accumulate at temperatures where 
the cooling curve $\Omega(T)$ has a positive slope.  This is a
manifestation of the Parker (1953) \& Field (1965) instability: when
$\frac{d\Omega(T)}{dT} > 0$, plasma cools more efficiently at slightly
higher temperatures and less efficiently at slightly lower
temperatures. At temperatures where
$\frac{d\Omega(T)}{dT} < 0$, plasma quickly cools to lower temperatures.
For solar-abundance plasmas, $\frac{d\Omega(T)}{dT} > 0$ in the ranges
0.7--1.2~MK, 4.5--7~MK, and above 21~MK. Thus, the shape of the cooling curve
explains the dip from 2--4~MK and from 9--20~MK seen in many EM distributions
and provides some physical justification for two- and three-temperature coronal
models.  The temperature of the hot component and the relative amount of
hot emission measure must reflect a
balance between magnetic heating and radiative cooling. For stars with
significant emission measure above 20~MK,
some mechanism must be super-heating the coronal plasma 
at fairly regular intervals.  

For three active, solar-mass stars observed with the {\it ASCA} SIS 
(EK~Dra, HN~Pegasii, and $\kappa^1$~Ceti in order of decreasing activity), 
G\"udel (1997) has modeled ${\rm Em}(T)$
by assuming X-rays are produced by an ensemble of flaring and cooling
loops. The G\"udel (1997) model is a modification of the solar
nanoflare model proposed by Kopp \& Poletto (1993).  
While the {\it ASCA} data cannot constrain many model parameters
(number of loops, loop dimension, and mean magnetic field strength),
the models can only reproduce the observed ${\rm Em}(T)$ provided that
the power-law distribution of flare
and microflare energies is $\alpha \approx 2$.
High S/N spectra of other active dwarfs
are needed to test the viability of continuous flaring and to establish
the importance of this mechanism as a function of rotation
rate and effective temperature.

In this paper, we present EUV and X-ray spectroscopy of HD~35850 =
HR~1817 (F8~V). HD~35850 was detected
serendipitously with {\it EXOSAT} (Cutispoto et al. 1991) and
follow-up optical spectroscopy has established it as a nearby, single,
solar-metallicity Pleiades-age ($t \approx 10^8$~yr), rapid rotator
(Tagliaferri et al. 1994).  
We are particularly interested in HD~35850 because it is
probably in the extreme state of magnetic activity for single,
main-sequence F stars.  There is considerable interest in probing the
magnetic dynamo in F stars because they possess relatively shallow
convection zones.  Kim \& Demarque estimate that,
in a $10^7$~yr-old $1.1$--$1.2M_{\odot}$ star like HD~35850,
the convective turnover time is $\tau_{\rm c} \approx 20$~d. 
In a young $1.0 M_{\odot}$ star
like EK~Dra, $\tau_{\rm c}\approx 40$~d.

\section{Observations and Data Analysis}

HD~35850 was observed by the {\it Extreme Ultraviolet Explorer}
from 1995 October 23 08:17:07 UT to 1995 October 30 07:20:34 UT.  
The {\it EUVE} Deep Survey/Spectrometer
consists of three aligned grazing-incidence telescopes.  The
telescope beams are intercepted by short-, medium-, and
long-wavelength reflection gratings and detected with
microchannel-plate detectors (SW, MW, and LW, respectively).   The
undispersed portions of the three beams are focussed onto the Deep
Survey microchannel-plate detector, DS.  This way, {\it EUVE} obtains
simultaneous, time-resolved 75--120~\AA\ broad-band photometry
and 70--160~\AA, 170--370~\AA, and 300--525~\AA\ medium-resolution spectra
(cf. Haisch, Bowyer, \& Malina 1993).
In support of this {\it EUVE} observing program,
\ion{Ca}{2} HK spectra were obtained on 1996 October 2--12.
HD~35850 was observed with the {\it ASCA} SIS and GIS for
$\sim 18$~ks on 1995 March 12. The reduction and analysis of these
data are described below.

\subsection{{\it EUVE} Deep Survey Data}

The {\it EUVE} Deep Survey/Spectrometer records photon events whenever the
source is visible by the satellite, including times when the satellite is
passing through the South Atlantic Anomaly (SAA).  These SAA passages are
characterized by increased background levels which lead to significant
and poorly characterized loss of telemetry, nicknamed ``primbsching".
Correcting for earth occultations,
the on-source time was $\sim 203$~ks.  Correcting for
instrument and telemetry dead times, the net exposure time was
$\sim 198$~ks.

\begin{figure}
\plotone{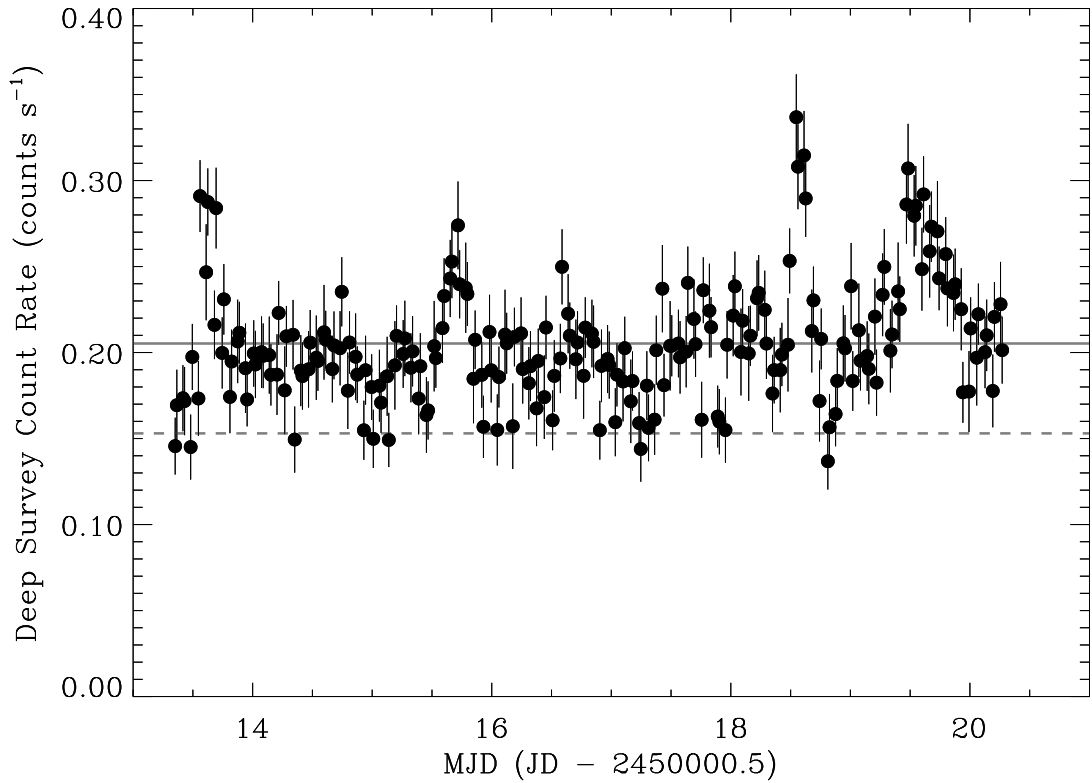}
\caption{EUVE Deep Survey light curve ($\sim 950$~s bins)
of HD~35850 obtained 1995 October 23--30.
The solid gray line indicates the mean DS count rate and the dashed gray line
shows the estimated activity level during the ASCA observation in 1995 March.}
\end{figure}

The Deep Survey (DS) events were screened at various count-rate
and primbsch-correction thresholds. Light curves were
generated with bin times ranging from 100 to 2000~s.  
Source events were extracted from a $25$~pixel radius circle centered 
on the PSF centroid.  
Background events were extracted from an annulus with inner
and outer radii of $37$ and $67$ pixels, respectively.
In Figure~1, we show the background-subtracted, dead-time and primbsch-corrected
75--120~\AA\ DS light curve for HD~35850 using 1000~s bins and
including times when primbsching (monitor Det7Q1dpc) was below 25\%.

The mean DS background-corrected count rate  of HD~35850 is
0.20~counts~s$^{-1}$.  No large flares are evident in the DS light
curve: the variability amplitude is approximately 100\% and the most
significant deviation from the mean count rate is a $5\sigma$ peak
near MJD 18.45.  The light curve, however, shows small to moderate
flares throughout the observation.  At least 7 events have a peak
count rate $3\sigma$ above the apparent ``quiescent'' level of
0.20~counts~s$^{-1}$.  To test the hypothesis that the observed counts
come from a non-variable  source, we have determined the cumulative
distribution function  of the unbinned photon arrival times for the
observed data and for simulated data from a constant,
0.2~counts~s$^{-1}$ source.   A two-sample Kolmogorov-Smirnov test
statistic (Feigelson \& Babu 1992) was calculated using the observed
and simulated distributions; the probability that the two data sets
were drawn from the same  parent population is low ($P \leq 10^{-9}$).
We conclude that the DS light curve exhibits significant variability. 

\begin{figure}
\plotone{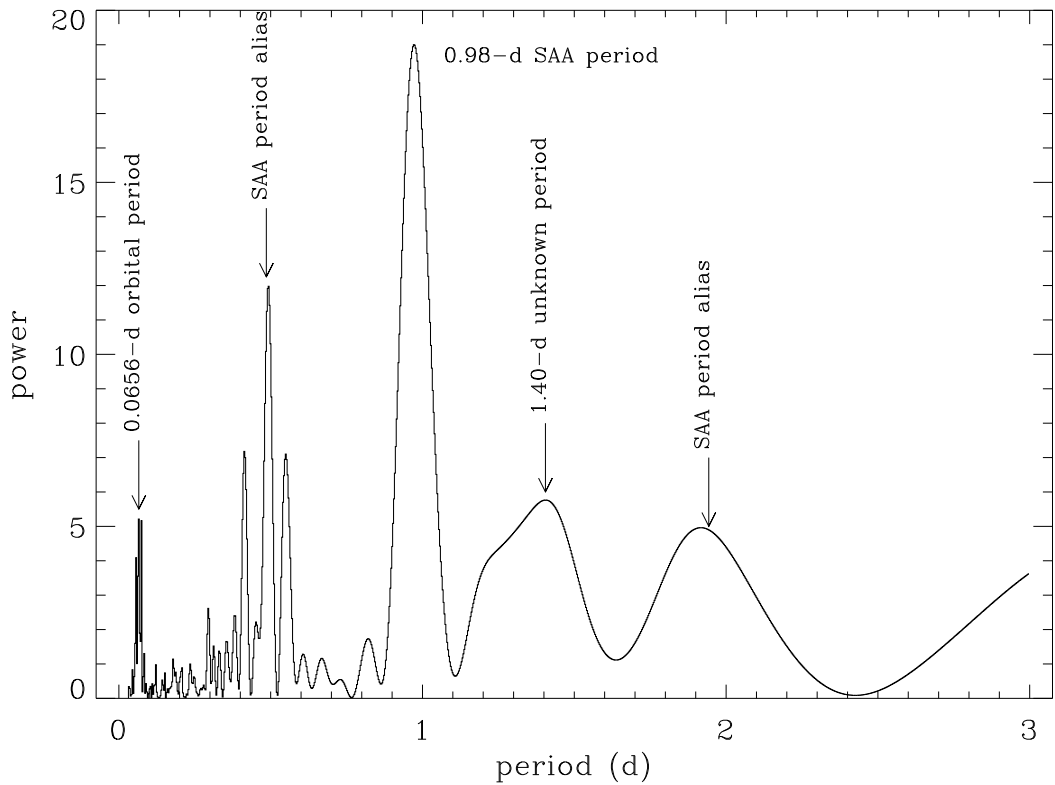}
\caption{
Scargle discrete Fourier transform for the binned DS light curve
in Fig.~1. The power spectrum indicates significant power at 0.0656~d,
0.98-d, and its aliases.  These periods are associated with satellite
SAA passages.  The only peak which cannot be assigned to systematic effects
occurs near 1.40~d.
}
\end{figure}

We have used the binned DS light curve to compute the discrete Fourier
transform (Scargle 1989), shown in Figure~2.  There is significant power 
around 0.98~d and its aliases.  For HD~35850,
$v\sin i \approx 50$~km~s$^{-1}$ and $R_{\star} \approx 1.18 R_{\odot}$
implies $P_{\rm rot} / \sin i \approx 1.1$~d, consistent with the 0.98-d period.
However, Halpern \& Marshall (1996) report that 0.98~d is a beat period
associated with the satellite's passage through the SAA.
For example, during the HD~35850 observation, the Det7Q1dpc monitor 
shows strong periodic signals at 0.984~d and at 0.0656~d.
Similar periods have been seen in other primbsch-corrected light curves
(Marshall 1998, private communication).
We thus conclude that some or all of the
0.98-d signal seen in the DS light curve results from SAA passages.  
We note that the only period seen in HD~35850's power spectrum that
{\em cannot} be attributed to primbsching is at
1.40~d.  Until further observations can be carried out, we tentatively
identify 1.40~d as the rotation period of HD~35850 (see \S 2.2).

\subsection{McMath-Pierce Optical Spectroscopy}

Medium-resolution \ion{Ca}{2} HK spectra were obtained with the
Solar-Stellar Spectrograph at the National Solar Observatory's
McMath-Pierce Telescope on 1996 October 2--12, ten days before the
{\it EUVE} observation.  The spectra around HK are typical of active,
late-type dwarfs, with broad absorption troughs and bright, narrow emission
cores.  To calculate the equivalent width of the H and K emission, 
we measured the excess emission in the line cores, normalizing by
the integrated continuum intensity in a 1~\AA\ box far from line center.

\begin{figure}
\plotone{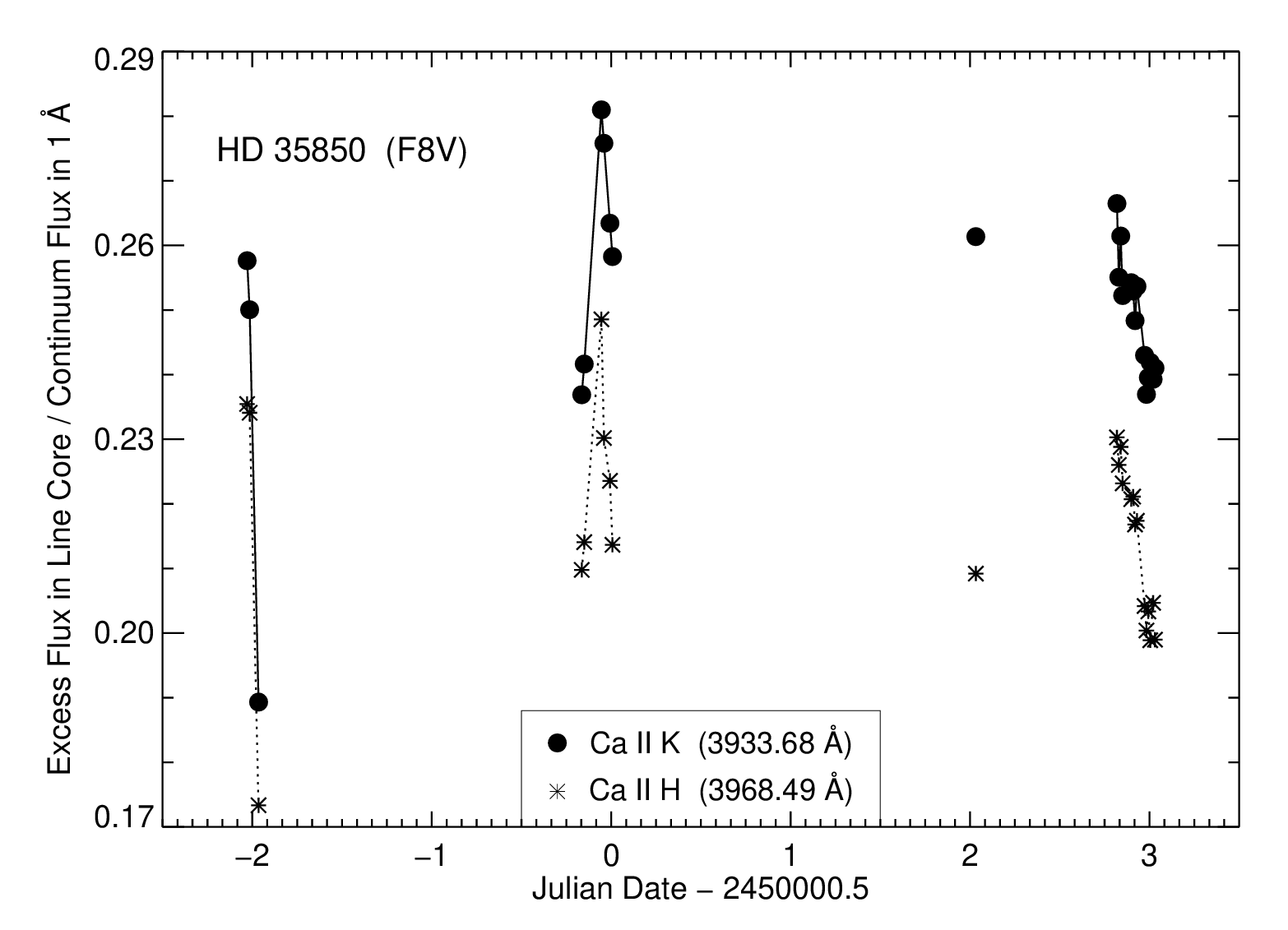}
\caption{
McMath-Pierce Solar-Stellar Spectrograph Ca~II HK light curves
of HD~35850 obtained 1995 October 2--12.  For each exposure, we determined
the excess emission in the line core in 1~\AA\ divided by the continuum flux
in 1~\AA\ (i.e., the equivalent emission width).
}
\end{figure}

In Figure~3 we plot the equivalent width of the 
\ion{Ca}{2} H (dots) and K (asterisks) emission cores.  While the
equivalent widths are highly variable on short time scales,
HD~35850 was only visible for a few hours each night.  The limited
sampling makes it difficult to estimate $P_{\rm rot}$ with these data
alone.  In Figure~4, we have phase-folded the summed H and K equivalent
widths using $P_{\rm rot} = 1.40$~d as derived from the DS photometry.  

\begin{figure}
\plotone{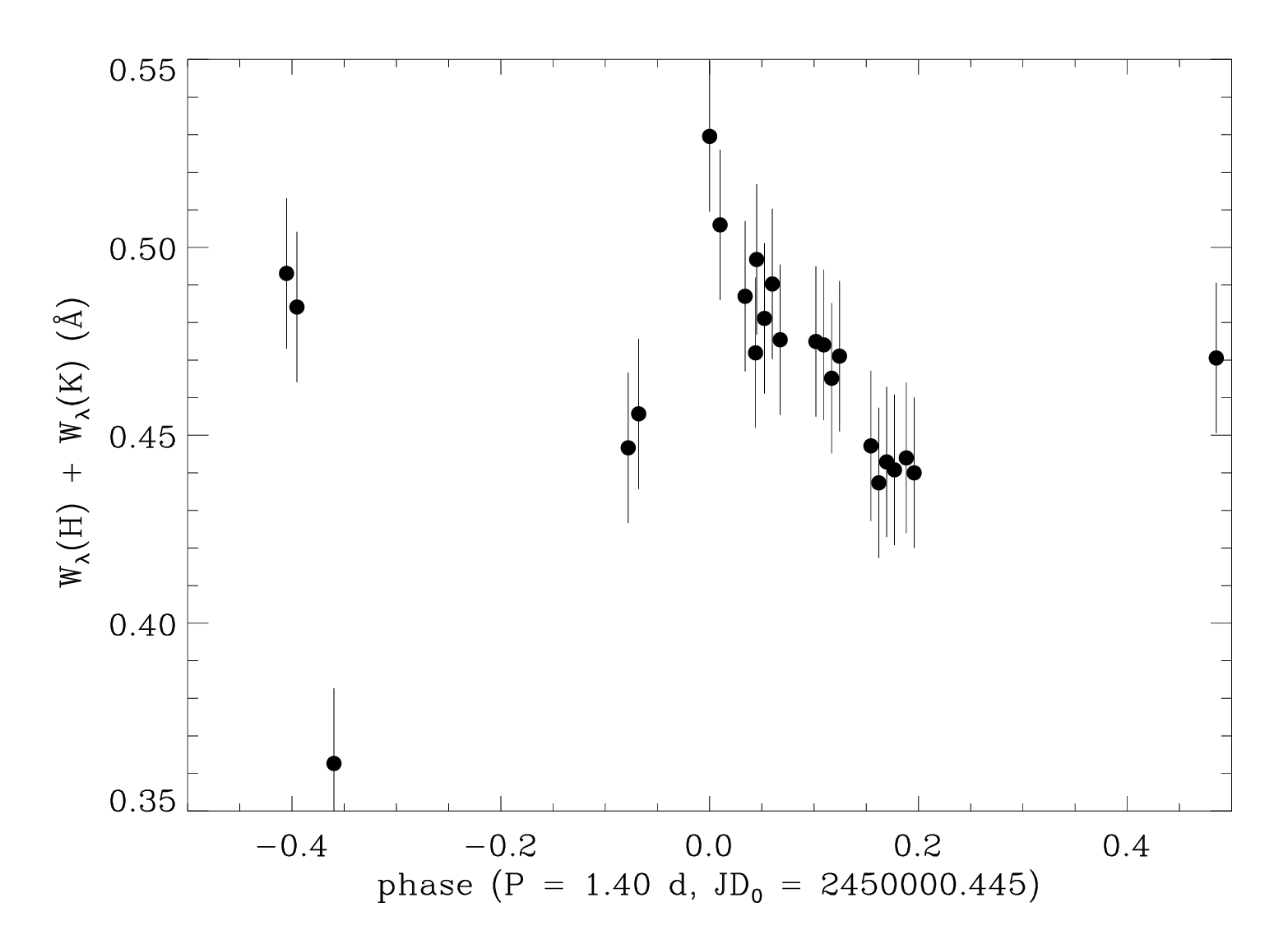}
\caption{
Phase-folded light curve of the combined Ca~II H and K equivalent
widths for a period of 1.405~d and zero-point of JD = 2450000.445, the time of
Ca~II HK emission maximum.
}
\end{figure}

\subsection{{\it EUVE} SW and MW Spectrometer Data}

For the SW, MW, and LW spectroscopic data, the photon event data were
screened to eliminate times of high background using the IRAF EUV package.
Specifically, data were excluded when the total detector count rate exceeded
40~counts~s$^{-1}$.  The SW, MW. and LW photon event lists were
used to create wavelength-calibrated FITS images and 
the SW, MW, and LW images were then analyzed within IDL.

We note that during the HD 35850
observation, the SW microchannel-plate detector developed a hot spot
8--12 detector pixels from the dispersion axis, close to the location
of the \ion{Fe}{22} 114~\AA\ line.  A light curve of the hot spot intensity
was used to identify the times when the hot spot flickered; those times were
removed from the photon event list.  Some hot pixels were still visible in the
SW image, so a $9\times 9$~pixel box around the hot spot was
set to the local background level.

Spectra have been extracted from the SW, MW, and LW
images using a modified version of the {\it IUE} SIPS
spectral extraction algorithm (Lenz \& Ayres 1992).
The optimal extraction routine is specifically tailored to low S/N
spectra from
photon-counting devices like the {\it EUVE} and {\it IUE}
spectrometers.  Briefly, a raw SW, MW, or LW $2048\times 2048$ pixel
image  is trimmed and compressed to $1024\times 290$ and 2-pixel
Gaussian smoothed.  Since the spectra are over-sampled (no less than 7
detector pixels per resolution element), 2-pixel compression and
smoothing improves S/N per bin without compromising spectral
resolution.  Using a predefined background region away from the source
spectrum chosen to reduce curved edge effects caused by the {\it EUVE}
reflection gratings, a background image is generated.  A
background-subtracted source image and corresponding error image are
then created.  The source image is used measure  the cross-dispersion
profile. To perform an optimal extraction, each row in the source
image is weighted by the cross-dispersion profile and the source
spectrum and corresponding error are extracted using a 7-pixel aperture.

\begin{figure}
\plotone{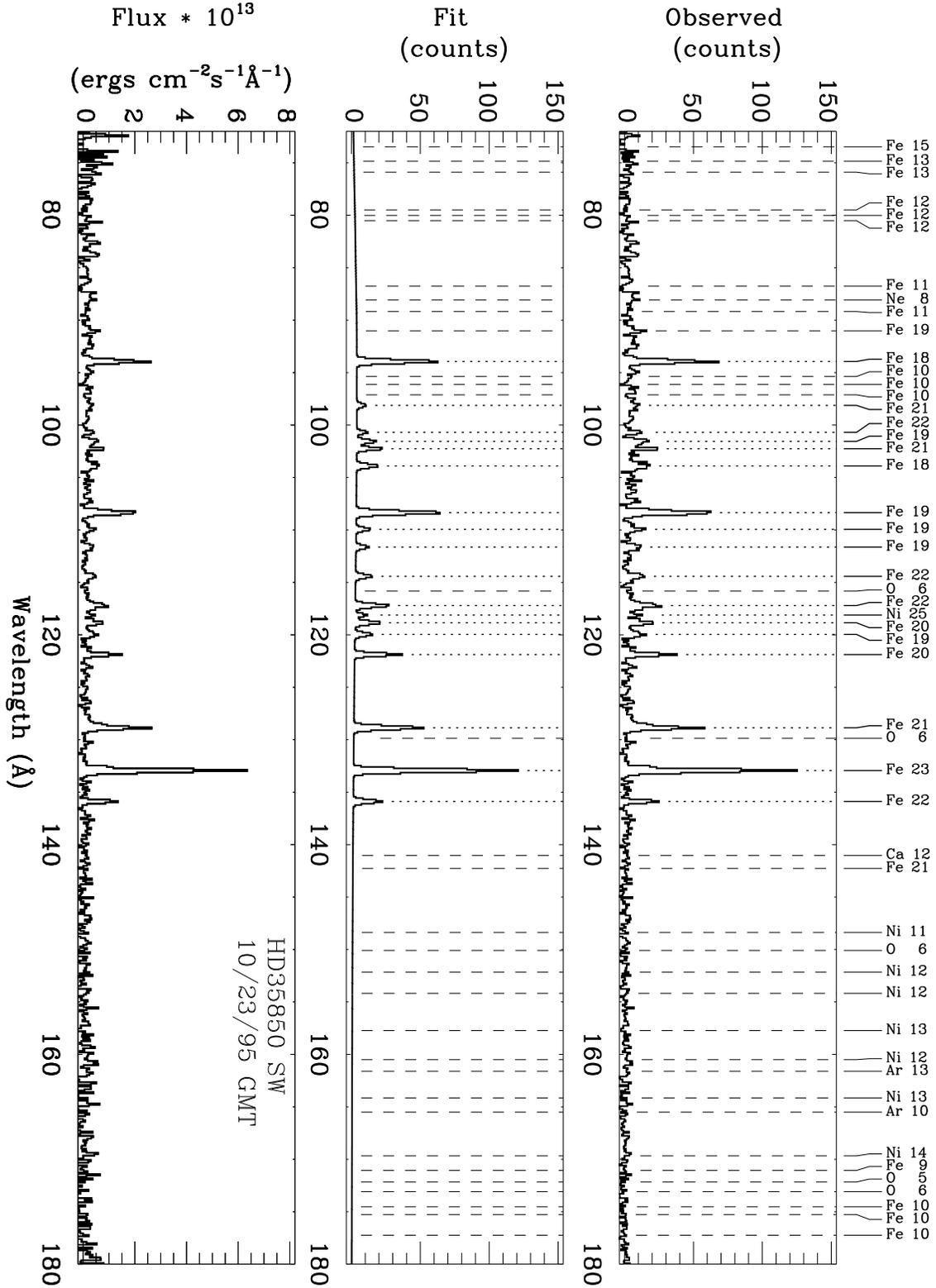}
\caption{
EUVE SW spectrum of HD~35850.
{\em Top panel}: observed count spectrum (0.135~\AA\ bins).
{\em Middle panel}: detected (dotted lines) and undetected (dashed lines) SW lines plus continuum.
}
\end{figure}

\begin{figure}
\plotone{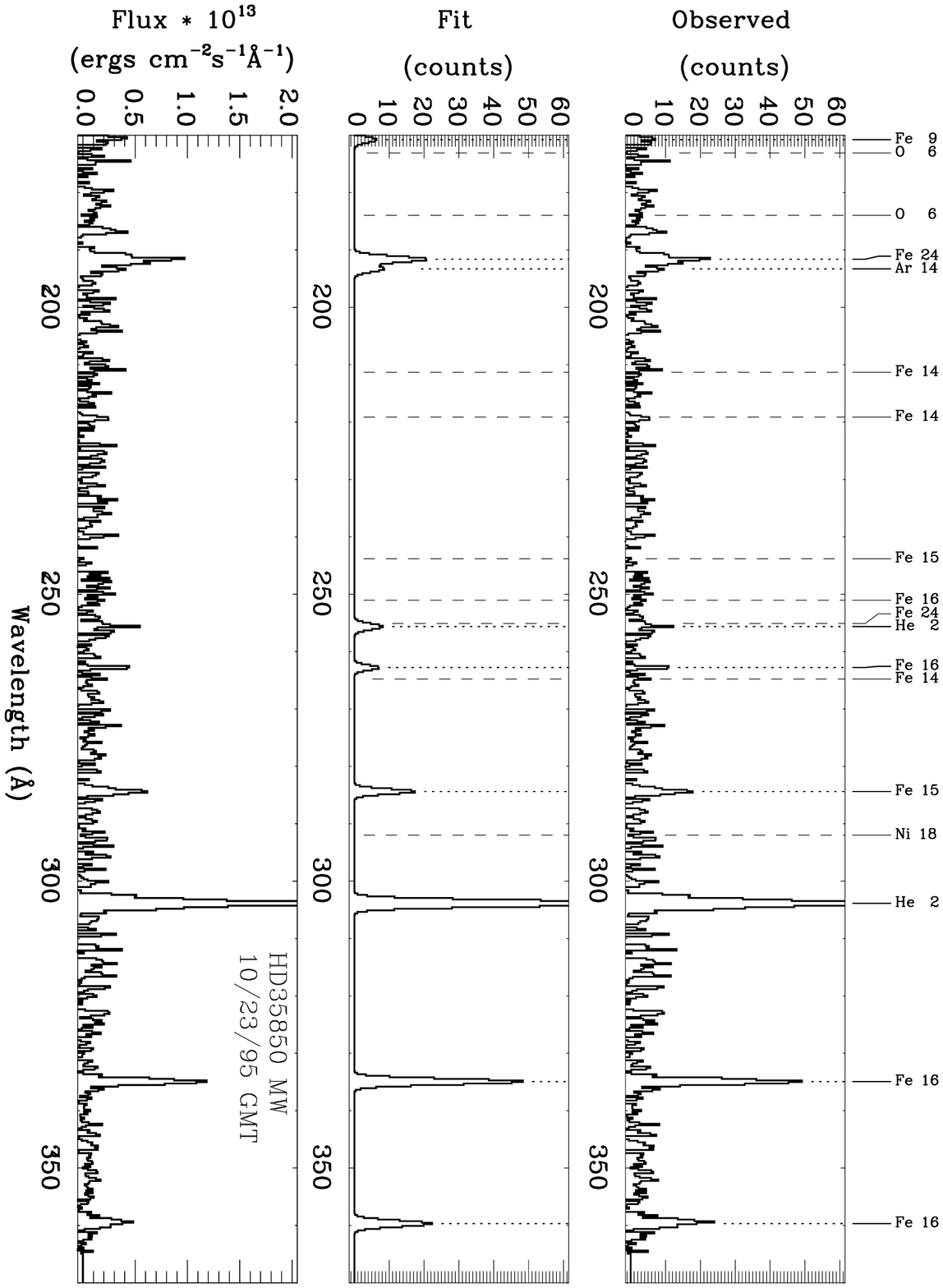}
\caption{
EUVE MW spectrum of HD~35850.
{\em Top panel}: observed count spectrum (0.270~\AA\ bins).
{\em Middle panel}: detected (dotted lines) and undetected (dashed lines) MW lines.
{\em Bottom panel}: flux-calibrated MW spectrum.
}
\end{figure}

No emission lines longward of 365~\AA\ were detected in the LW count spectrum,
presumably because of increased interstellar absorption at long wavelengths.  
Since the lines shortward of 365~\AA\ are measured in the MW spectrum,
we did not consider the LW spectrum in our analysis.  In Figures~5
and 6 (bottom panels), we show the flux-calibrated SW and MW spectra.

\subsubsection{IDL Line and Continuum Fitting}

In order to measure line fluxes, we have constructed a line list using
the emissivity lists of Brickhouse et al. (1995),
Monsigniori Fossi \& Landini (1993), and Mewe et al. (1985), in order
of priority.  Lines were grouped to match the SW and MW spectral resolution
(0.5 and 1~\AA, respectively), allowing us to account for emissivity from
blended lines.  Based on the ionization states of the brightest
identified lines, e.g., \ion{Fe}{15}, \ion{Fe}{16}, and \ion{Fe}{20},
we made a starting guess at the plasma temperature ($\log T \approx 6.8$)
and assumed a density $n_{\rm e} = 10^{11}$~cm$^{-3}$.
The line emissivities, spectrometer effective area curves, and
exposure time were used to provide a starting guess for each line
strength.  The observed spectrum was then fit with multiple Gaussians,
varying wavelength ($\pm$ one resolution element) and line strength.
Lines with fewer than $2\sigma$ counts were eliminated from the fit
and the procedure was repeated.  This way, line counts and upper
limits were determined for all high-emissivity \ion{Fe}{9} to \ion{Fe}{24}
lines.

The SW spectrum shows a small but detectable thermal bremsstrahlung
continuum.  To determine the SW continuum luminosity from 85--135~\AA,
the continuum bins were fit with two parameters (emission measure
and temperature) using the continuum emissivity model of Mewe et
al. (1985).  The Hipparcos distance to HD~35850 is 26.8~pc
(Perryman et al. 1997).  For HD~35850, the 85--135~\AA\ continuum luminosity
is $L_{\rm cont} \approx 4.6\pm0.4\times 10^{28}$~ergs~s$^{-1}$.

In Figs.~5 and 6 (top and middle panels), we show the observed and
fit count spectra with line identifications.  Detected lines are
indicated with dotted lines and upper limits are indicated with dashed
lines. In Tables~1 and 2, we list
the ion, possible blends, temperature of peak emissivity, 
laboratory and measured wavelengths, a detection flag,
measured line luminosities, and the signal-to-noise ratio for detected lines.
The line luminosities in Tables 1 and 2 use
$N_{\rm H} = 1.7 \times 10^{18}$~cm$^{-2}$ (see \S 2.3.3).
In all, 28 distinct lines are detected above
$2\sigma$ in the SW and MW spectra.

\subsubsection{Electron-Density and Column-Density Estimates}

The \ion{Fe}{15} and \ion{Fe}{16}  $\lambda\lambda 285, 335, 365$ line
ratios are relatively insensitive to density and provide the best
estimate of interstellar $N_{\rm H}$ in the {\it EUVE} bandpass.  We find
$N_{\rm H} = 1.9\pm 0.4 \times 10^{18}$~cm$^{-2}$.  

\begin{figure}
\plotone{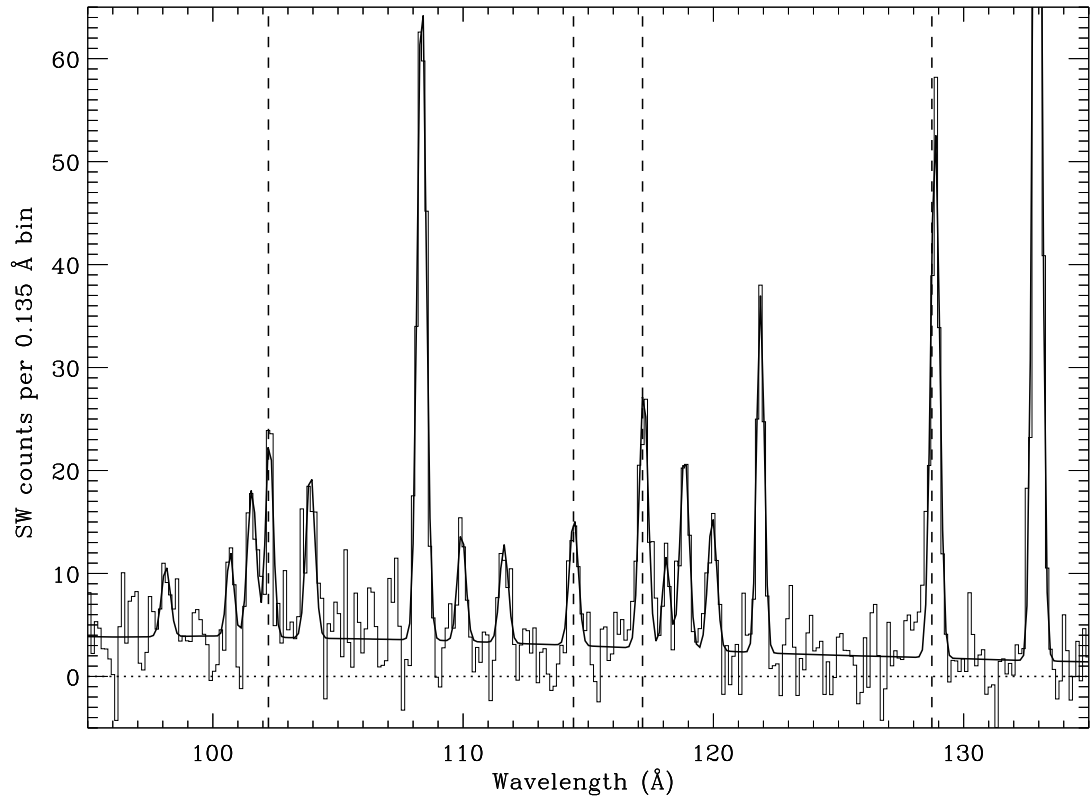}
\caption{
SW spectrum of HD~35850 in the 95--135~\AA\ region (histogram) and our fit
to the lines and continuum (solid curve).  The dashed lines indicate the
wavelength of the density-sensitive Fe~XXI and Fe~XXII lines.}
\end{figure}

Two density-sensitive line ratios can be measured accurately from the SW
spectrum of HD 35850: \ion{Fe}{22} $\lambda\lambda 114, 117$ 
and \ion{Fe}{21} $\lambda\lambda 102, 129$. In Figure~7 we
plot the observed SW spectrum (histogram) and fitted lines and continuum
(solid line).  The density-sensitive lines are indicated with dashed lines.
The line \ion{Fe}{21} line ratio is $0.22 \pm 0.05$.
The predicted branching ratio (solid curve)
and observed line ratio (cross with 1$\sigma$ error bar) are plotted
versus density in Figure~8.
Fig.~8 suggests that $\log n_{\rm e} < 11.6$~cm$^{-3}$.  
Consequently, we have chosen
$\log n_{\rm e} = 11.0$~cm$^{-3}$ for the IDL and SPEX
emission-measure analyses.

\begin{figure}
\plotone{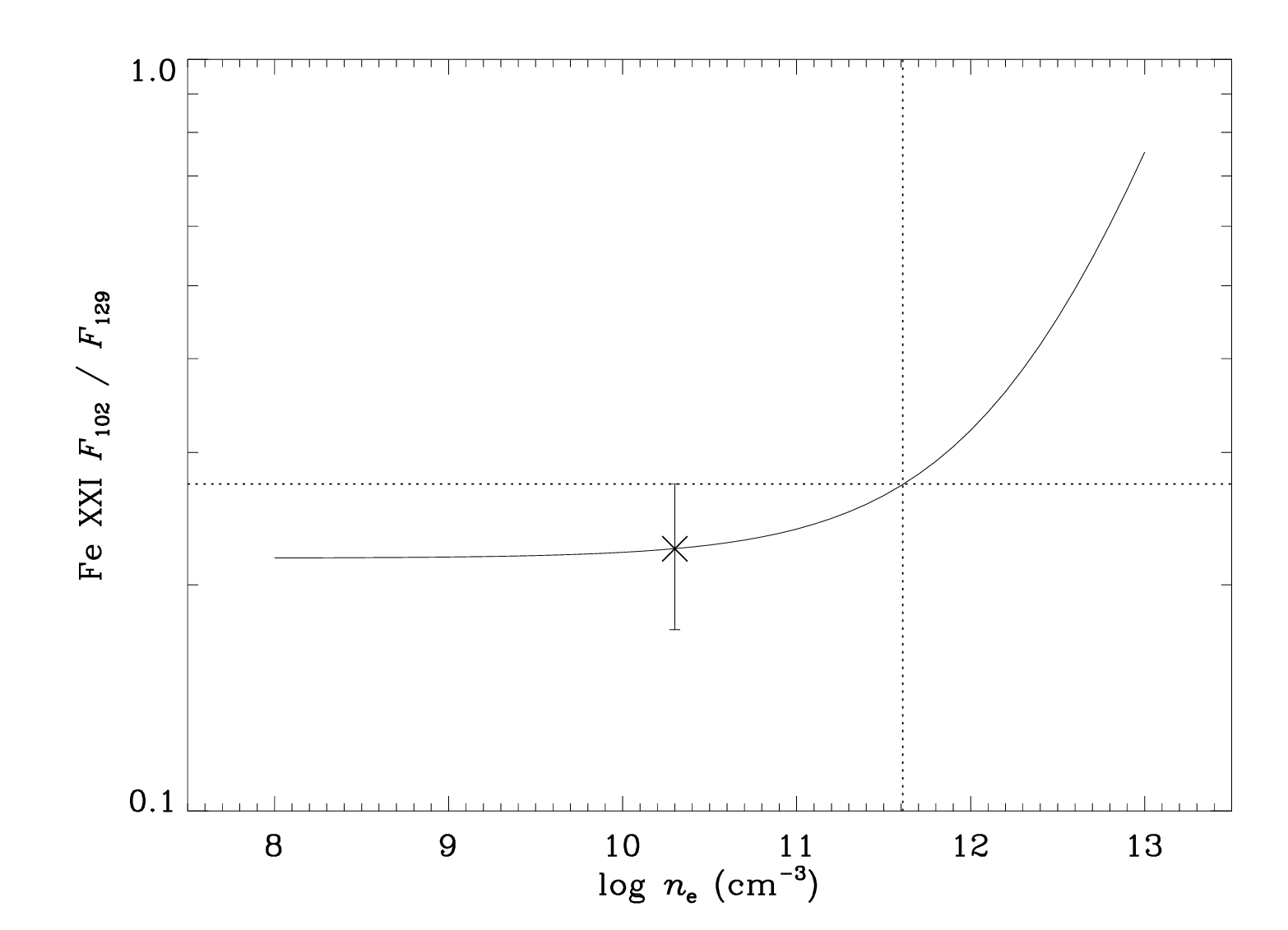}
\caption{Fe~XXI branching ratio as a function of $\log n_{\rm e}$.  The measured
Fe~XXI $\lambda\lambda 102, 129$ ratio and $1\sigma$ error are plotted.  Based
on the predicted branching ratio, we derive an upper limit
$\log n_{\rm e} < 11.6$~cm$^{-3}$.}
\end{figure}

The \ion{Fe}{22} line ratio is $0.47 \pm 0.14$, suggesting
$n_{\rm e} > 10^{13}$~cm$^{-3}$.  This ratio is unusually
high and the implied density does not agree with the more reliable
\ion{Fe}{21} measurement.
The detector hot spot (\S~2.3) may have contributed to
the anomalous 114~\AA\ flux.  
We note that line ratio calculations may have systematic
uncertainties of up to 50\% (Brickhouse et al. 1995) and that the
derived density limit is only approximate.

\subsubsection{IDL Emission-Measure Analysis}

The measured line luminosities and upper limits have been used
to estimate the coronal emission-measure distribution, ${\rm Em}(T)$.  
Previous analyses of {\it EUVE} spectra (e.g., G\"udel et al. 1997;
Mewe et al. 1996) 
have generally relied on global fits to estimate ${\rm Em}(T)$.
For low S/N spectra like these, we prefer to use the
bright lines in Tables~1 and 2 because
(i) the emissivities of brighter lines have been fairly well established and
(ii) the $\chi^2$ statistic is not dominated by continuum bins.

We have excluded the \ion{He}{2} $\lambda 304$  line from our
analysis.  On the Sun, the \ion{He}{2} Ly$\alpha$ line is not
a reliable temperature indicator: some, perhaps most, of the \ion{He}{2}
emission occurs in the chromosphere as a result of back-heating from the
corona. 
We note also that around $\lambda 255$, the \ion{Fe}{24} line is blended
with \ion{He}{2}.  The combined flux from both lines is listed as an upper
limit in Table~2.  For the remaining lines, our fitting procedure minimizes
$\chi^2$ between the measured and predicted line luminosities by varying
the amplitude and shape of ${\rm Em}(\log T)$ (in $\Delta \log T = 0.1$ bins).

We use exponential Chebyshev polynomials (Lemen et al. 1989)
to describe ${\rm Em}(\log T)$.
In addition to detected emission
lines, the undetected \ion{Fe}{10} to \ion{Fe}{14} lines 
have been used as additional constraints
in the non-linear least-squares fitting.  The SW continuum 
luminosity serves to constrain Fe abundance.

Best-fit emission-measure distributions have been derived for a grid of
column densities, Fe abundances, and Chebyshev-polynomial orders:
$0.0 \leq N_{\rm H} \leq 4.0\times 10^{18}$~cm$^{-2}$, 
$0.1 \leq Z \leq 2.0$, and $2 \leq n \leq 11$, 
where $Z$ is the coronal Fe abundance relative
to the solar photospheric value of Anders \& Grevesse (1989).
We obtain acceptable $\chi_{\nu}^2$ values for all $n \geq 6$
and the resulting ${\rm Em}(T)$ are characterized by peaks at
$\log T \approx 6.8$ and $\log T \approx 7.4$.  
The best-fit column densities, abundances, and emission-measure
distributions are not a strong function of $n$.  As a result, 
we choose $n=6$, the lowest polynomial order for which we obtain good fits.

Fits using different order polynomials orders, column densities,
and abundances yield qualitatively similar results: prominent peaks
in the emission-measure distribution at $\log T \approx 6.8$ and
$\log T \approx 7.4$, although the sharpness of the high-T peak is not
well constrained.  The relative strength of the high-T component increases 
with increasing abundance.  The best-fit emission-measure
distributions all go to zero at very high temperatures ($\log T > 7.7$).  

\begin{figure}
\plotone{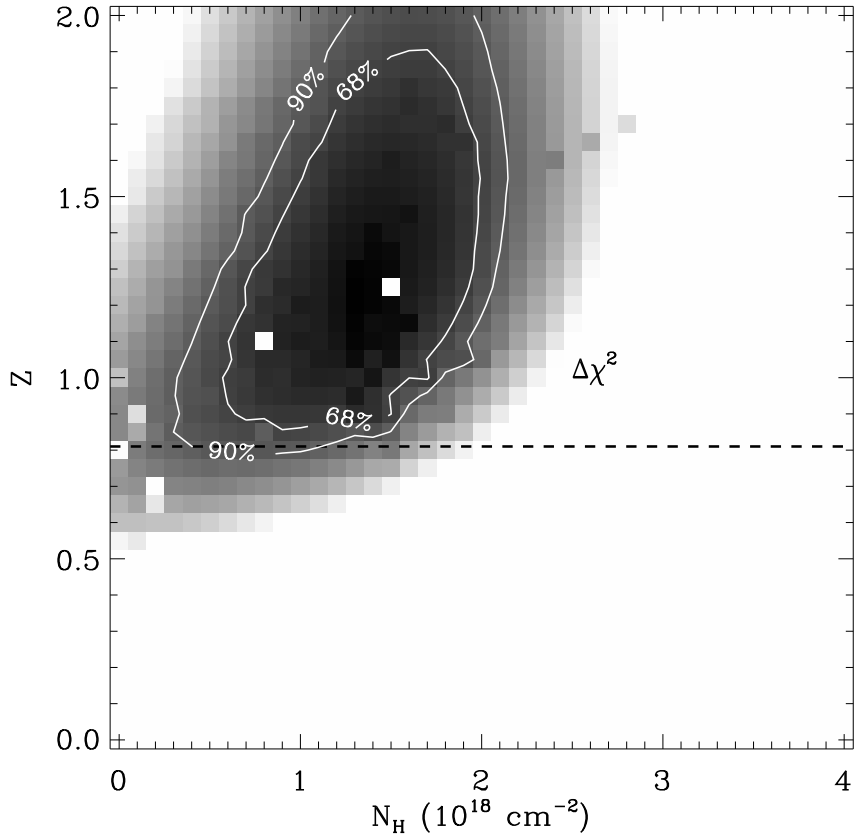}
\caption{Grid of $\log \Delta\chi^2$.  The SW and MW lines and continuum
luminosities were fit using 8th-order Chebyshev polynomials to describe
${\rm Em}(\log T)$ at each point in the $Z$ versus $N_{\rm H}$ plane,
where $Z$ is the coronal Fe abundance relative to the solar photospheric values
of Anders \& Grevesse (1989).
Dark pixels indicate regions of low $\chi^2$.  The 68 and 90\% confidence
contours for 9 free parameters are shown.}
\end{figure}

In Figure~9, we plot $\log \Delta\chi^2$ as a function of $N_{\rm H}$ and
$Z$.  The best fit ($\chi_{\nu}^2 \approx 1.07$) is obtained for
$N_{\rm H} = 1.7 \times 10^{18}$~cm$^{-2}$ and $Z=1.15$.  
The 68\% and 90\% confidence contours are plotted for
9 free parameters ($Z$, $N_{\rm H}$, and $n = 0, \dots, 6$).  
The {\it EUVE} line-to-continuum ratio yields $Z = 1.15_{-0.35}^{+0.75}$.
To obtain substantially sub-solar abundances ($Z < 0.5$), would require
more than doubling the continuum level (dark solid line in Fig.~7).  
This is not compatible with the background-subtracted SW spectrum.
We have experimented with various background-subtraction algorithms (smoothed,
unsmoothed, polynomial fit): all produce bright lines and a weak continuum.

\begin{figure}
\plotone{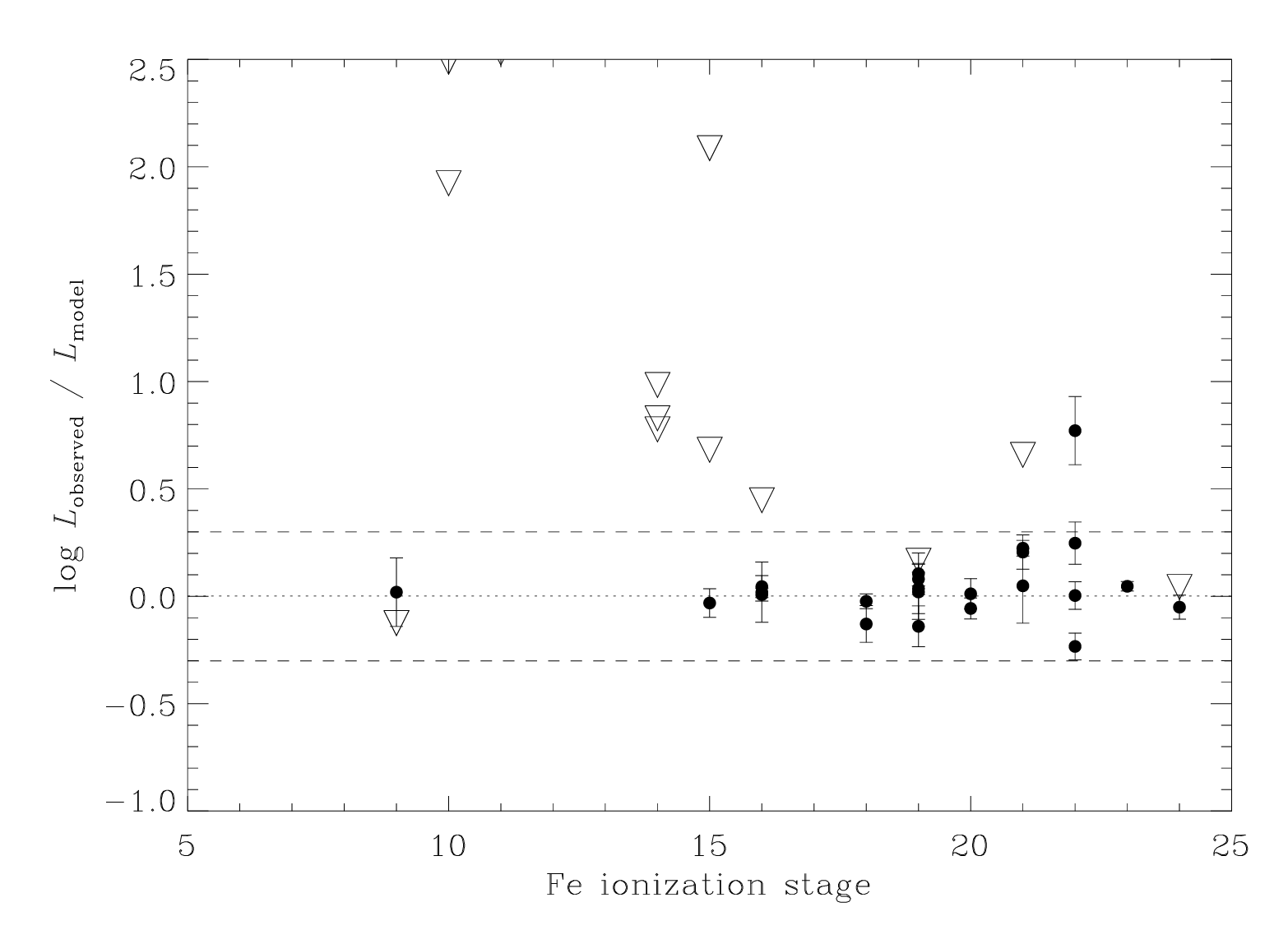}
\caption{Comparison of observed to predicted line luminosities for a range of Fe
ionization states for $Z = 1$, $N_{\rm H} = 1.4\times 10^{18}$~cm$^{-2}$, and
$\log n_{\rm e}=11.0$~cm$^{-3}$. Downward-pointing triangles indicate upper
limits for high-emissivity, undetected lines. Note the discrepant Fe~XXII
measurement.}
\end{figure}

We find adequate fits for a range of column densities and abundances, and,
in particular, we find $\chi_{\nu}^2 \approx 1.15$ for
$N_{\rm H} = 1.4\times 10^{18}$~cm$^{-2}$ and $Z = 1$.
We adopt these values in Figs.~10--14.  In Figure~10,
the observed and predicted line luminosities for detected
(filled circles) and undetected (open triangles) lines are compared,
with a factor of two deviation indicated with dashed lines.
Note the anomalously high \ion{Fe}{22} line flux (see \S 2.3.2).
In Figure~11, we plot ${\rm Em}(\log T)$ per $\log T$ bin of 0.1 dex
as derived from the IDL line analysis (dash-dotted line).

\begin{figure}
\plotone{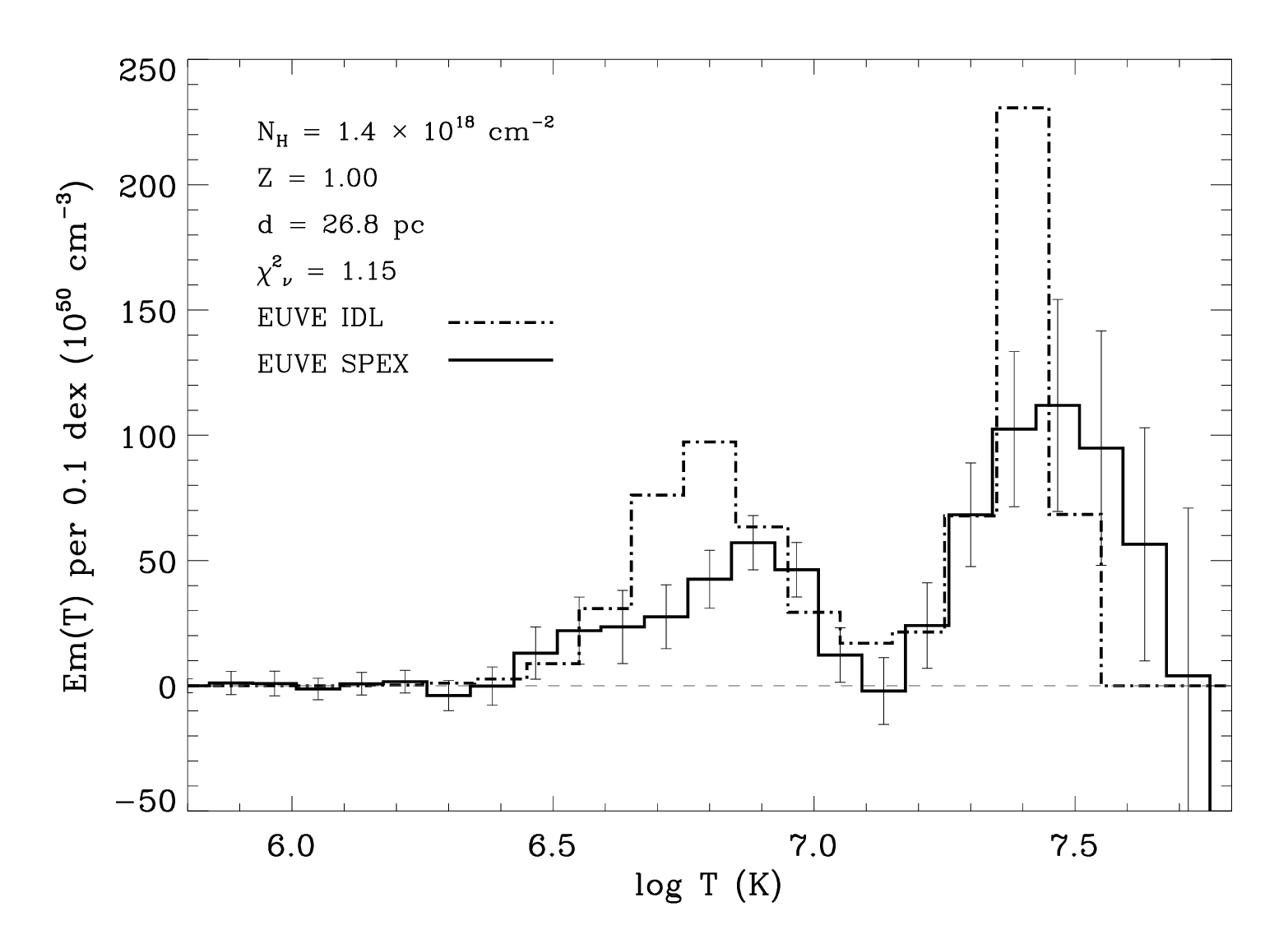}
\caption{Best-fit emission-measure distribution based on
the IDL (dash-dotted line) and SPEX (solid line and error bars) analyses
of the EUVE SW and MW lines and continuum.}
\end{figure}

\subsubsection{SPEX Differential Emission-Measure Analysis}

In order to verify the results of the IDL line analysis, we performed
a differential emission measure analysis of the optimally-extracted
{\it EUVE} SW and MW spectra using the SPEX code
(see Kaastra et al. 1992 for a detailed description of the DEM
method and the SPEX Collisional Ionization Equilibrium model).  
For fixed values of $N_{\rm H}$ and $Z$, we
derived ${\rm Em}(T)$ using the SPEX regularization method.
In Fig.~11, we plot ${\rm Em}(T)$ for 
$N_{\rm H}=1.4\times 10^{18}$~cm$^{-2}$ and $Z = 1$ (solid line).
The IDL and SPEX analyses yield similar emission-measure distributions,
with 55--60\% of the emission measure in the hotter component 
above $\log T = 7$.

\subsection{{\it ASCA} SIS data}

An 18-ks {\it ASCA} exposure of HD~35850 obtained in 1995 has been analyzed
by Tagliaferri et al. (1997).  We use their reduced {\it ASCA} SIS0 spectrum
to further constrain the {\it EUVE} results.
Because the SIS cannot resolve individual emission lines from coronal
sources, we cannot use the IDL line analysis method described in
\S~2.1.4.  Consequently, we fitted the {\it ASCA} SIS0 spectrum using
the SPEX DEM code.  For the SIS0 fits, $N_{\rm H}$ was fixed at
$1.4 \times 10^{18}$~cm$^{-2}$. 
The best-fit abundance using the DEM collisional ionization equilibrium (CIE)
model is $Z \approx 0.5$ with acceptable fits in the range
0.34--0.81. The {\it ASCA} upper bound on $Z$ is plotted in Fig.~9
as a dashed line.
We note that
$Z \approx 0.8$ is marginally compatible with {\it EUVE} and
{\it ASCA}.

\begin{figure}
\plotone{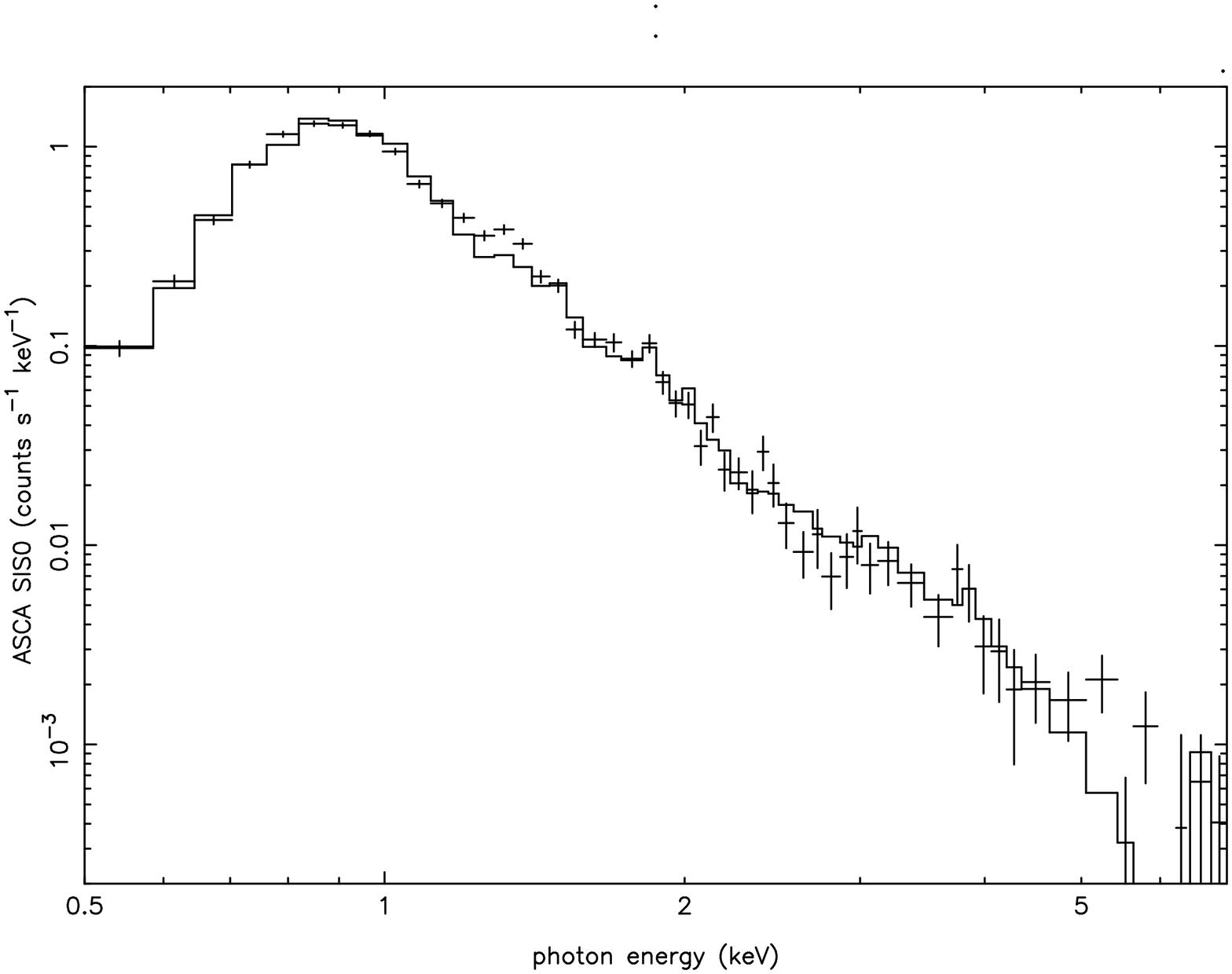}
\caption{The 18~ks ASCA SIS0 spectrum of HD~35850 obtained 1995 March 12
(points and error bars).  The best-fit SPEX DEM model (not shown) indicates
sub-solar Fe abundances, $0.34 < $Z$ < 0.81$.  The model shown (solid histogram)
assumes $Z = 1$, $N_{\rm H} = 1.4\times 10^{18}$~cm$^{-2}$, and
$\log n_{\rm e} = 11.0$~cm$^{-3}$.  The SIS0 spectrum is generally consistent
with solar abundances aside from serious departures around 1.2~keV and 2.4~keV.}
\end{figure}

To illustrate, the {\it ASCA} SIS0 data are shown in Figure~12 (points)
with the DEM model spectrum (histogram) for $Z = 1$.
Discrepancies between the SIS data and the $Z = 1$
model are seen around 1.2~keV and 2.4~keV.
These are the spectral features which normally 
drive the fit towards lower abundances.
Brickhouse et al. (1997) have identified a complex of lines from
1.0--1.3~keV from highly excited ($n > 5$), highly ionized
(\ion{Fe}{17} to \ion{Fe}{25}) states that are missing from the
plasma codes in XSPEC and SPEX.  The missing lines
may partly explain the large discrepancy around 1.2~keV.

\begin{figure}
\plotone{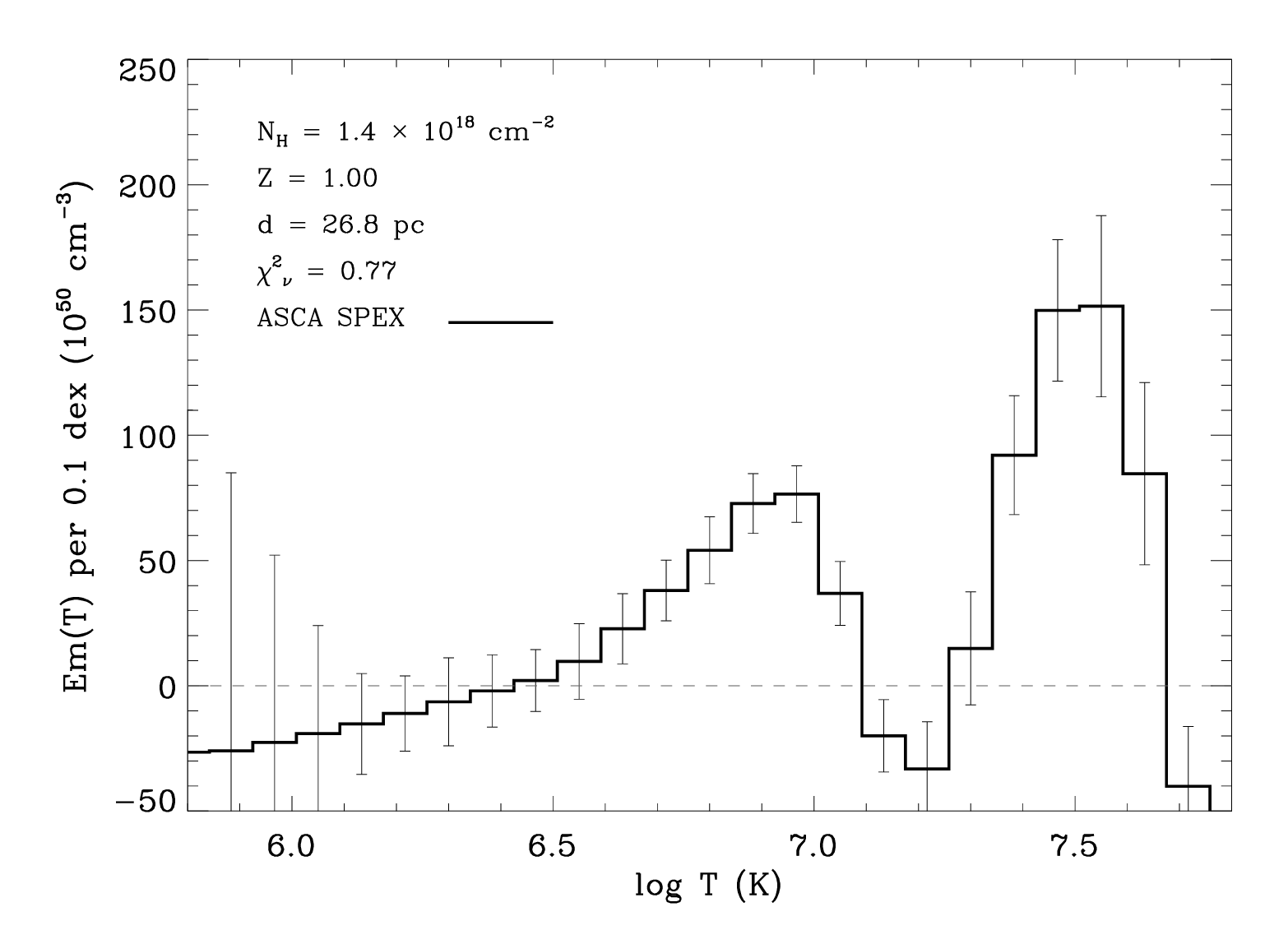}
\caption{Emission-measure distribution based on the SPEX global-fitting
of the ASCA SIS0 spectrum.  The EUVE and ASCA data
were not obtained simultaneously.  Poor fits resulted when both data sets
were modeled together.}
\end{figure}

In Figure~13, we show the corresponding ASCA SPEX
${\rm Em}(\log T)$ for $Z = 1$.
Like the {\it EUVE} emission-measure distributions,
the {\it ASCA} ${\rm Em}(T)$ peaks at $\log T$ of 6.7 and 7.4.  
Finally, we note that the {\it EUVE} count rates are higher than expected
from the SIS spectrum.
Based on the SIS emission-measure analysis,
the expected {\it EUVE} DS count rate is approximately
0.16~counts~s$^{-1}$, indicated in Fig.~1 as a dashed line.
HD~35850 appears to have been more active (by $\sim 25\%$)
during the {\it EUVE} observation.   

\subsection{Comparison with Previous Results}
 
For HD~35850, we find that the {\it EUVE} line-to-continuum ratio
indicates approximately solar coronal Fe abundance while a SPEX DEM analysis
of the {\it ASCA} SIS0 spectrum is consistent with moderately
sub-solar abundances.  We note that HD~35850's photospheric Fe abundance
is close to solar (Tagliaferri et al. 1994).  Analyses of other active
coronal sources often find coronal abundances far below the measured
photospheric abundances (e.g., AB~Dor, Mewe et al. 1996).
 
The {\it ASCA} SIS and GIS spectra of HD~35850 were fit by Tagliaferri et al.
(1997) using a number of multi-temperature plasma models in XSPEC (Arnaud 1996).
For example, the best-fit SIS0 MEKAL parameters indicate a
two-temperature corona ($kT_1 \approx 0.6$~keV and $kT_2 \approx 1.2$~keV)
with sub-solar abundances ($0.12 < Z < 0.25$).
Tagliaferri et al. (1997) also performed a combined PSPC/SIS/GIS
analysis using a 5-ks {\it ROSAT} PSPC observation of a field near HD~35850.
The combined analysis involved comparing 2-T and 3-T MEKAL and RS models
in XSPEC with solar and non-solar abundances and using different
cross-detector normalization constraints (see their Table~5 and Figure~3).
The XSPEC fits generally favor sub-solar abundance models; the best-fit
MEKAL 3-T model yields $Z = 0.34 \pm 0.04$, $T_1 = 0.52\pm0.07$,
$T_2 = 0.78^{+0.15}_{-0.09}$, and $T_3 = 1.9^{+1.0}_{-0.4}$, with
$\approx 47\%$ of the total emission measure in the coolest component.  This
cool component is the $\log T = 6.8$ component seen in the {\it EUVE} DEM.
The two hottest components appear to represent the hotter DEM component.
Our DEM analysis of the ASCA SIS spectra yields somewhat higher abundances
($0.34 < Z < 0.81$) than Tagliaferri et al. (1997) and somewhat lower
Fe abundance than our {\it EUVE} analysis.
 
Systematic differences between {\it ASCA} and
{\it EUVE} analyses have been reported for other bright coronal sources.
For example, Brickhouse et al. (1997) obtained simultaneous observations
of Capella with {\it ASCA} and {\it EUVE} in 1996 March.
As with previously obtained {\it EUVE}
spectra of Capella, they find essentially solar photospheric
Fe abundance.  On the other hand, the plasma codes in XSPEC yield
consistently poorer fits to the {\it ASCA} SIS spectra of Capella and
require sub-solar abundances ($Z \approx 0.7$).
 
{\em Note:} After the submission of this paper, a preprint of a paper to
appear in {\it Astronomy and Astrophysics} by Mathioudakis \& Mullan (1998)
came to our attention which analyzes the {\it EUVE} observations of
HR~1817=HD~35850.  This paper and the Mathioudakis \& Mullan
paper arrive at substantially different conclusions regarding the Fe abundance
of HD~35850.  Mathioudakis \& Mullan (1998) visually compare the
observed {\it EUVE} SW and MW spectra with synthetic spectra based
on the 3-T MEKAL model suggested by Tagliaferri et al. (1997).
They conclude that the {\it EUVE} spectra are consistent with 
Tagliaferri et al. (1997) and are
inconsistent with any solar-abundance model, particularly because the 
\ion{Fe}{13}--\ion{Fe}{15} lines in the MW spectrum are so weak.
While it is difficult to visually compare spectra, their observed and
synthetic MW spectra appear to be quite different: the predicted lines
are weak and the the low-Z model appears to over-predict the MW continuum 
(see Figures~2 and 5 in Mathioudakis \& Mullan 1998).
Given the limitations of {\it ASCA} and {\it EUVE},
higher-resolution, higher-S/N {\it AXAF} HETG or {\it XMM} RGS
spectra may be needed to determine HD~35850's coronal abundance.

\section{Microflaring}

The double-peaked EM distribution observed on three active,
solar-type stars has been modeled by G\"udel (1997)
on the basis of a simplified stochastic flare model derived from
the solar nanoflare model of Kopp \& Poletto (1993). 
Rather than approaching the EM distribution problem hydrostatically,
as is the case when using loop scaling laws, Kopp \& Poletto (1993)
use a simplified hydrodynamic model to treat a large number of flares.

The flare model reduces the loop hydrodynamics
to a point model with a chromospheric energy sink.  A series
of energy pulses of finite duration is fed into the loop.  
This heating energy is lost by conduction into the chromosphere
and by radiation into space.
Without flaring, all loops are kept at the same equilibrium temperature 
dictated by hydrostatic loop scaling laws (Rosner, Tucker, \& Vaiana 1978). 
Therefore, our model starts out at a lower threshold 
temperature and does not consider cooler loops. 
The salient feature of this model is its phenomenological
similarity with more sophisticated hydrodynamic simulations in terms of the
emission measure, temperature, and radiation history.
The point model cannot treat the
more complicated 1-D problem of radiation and conduction 
occurring between the corona, the transition region, and the chromosphere.
The need to simulate a large number
of flares requires such a simplified approach.

Flares are ignited in each loop randomly distributed in time but satisfying
a statistical number distribution in total flare energy $E$ above
a pre-defined threshold energy. I.e.,
${\rm d}N/{\rm d}E \propto E^{-\alpha}$ where $N(E)$ is the 
number density of flares in the energy interval $[E, E+{\rm d}E]$.
On the basis of solar and stellar optical flare monitoring,
$1.8 < \alpha < 2.0$ has been suggested by Hudson (1991).
For consistency with G\"udel (1997),
we chose a threshold energy $E > 10^{27}$~ergs. In order to
reduce the number of free parameters, all loops have the same height and
vary only in their thickness between a minimum of $6.4\times 10^6$~cm
and a maximum of $4.3\times 10^9$~cm to accommodate the
range of flare energies. Each run produces a large number of small flares
and only a small number of very large flares.
Because no very large flares were
observed by {\it EUVE}, the largest flares in the simulations were
eliminated.   

The total EM distribution is determined by averaging (in time and space)
the EM distribution of all loops. Similarly,
the light curve is the superposition of coronal radiation leaving
the star into $2\pi$ steradians, i.e., half the radiative losses are
absorbed in the chromosphere, half are radiated into space.
We have experimented with various parameters
and find that our model reproduces the required coronal 
luminosity and EM distribution only if
we assume compact loops, with a semi-length of 
$2.6\times 10^9$~cm and essentially full surface coverage.
The simulations provide fairly tight constraints on the power-lax
index of flare energies ($\alpha \approx 1.8$).
In this case, the equilibrium temperature is $6\times 10^6$~K.

\begin{figure}
\plotone{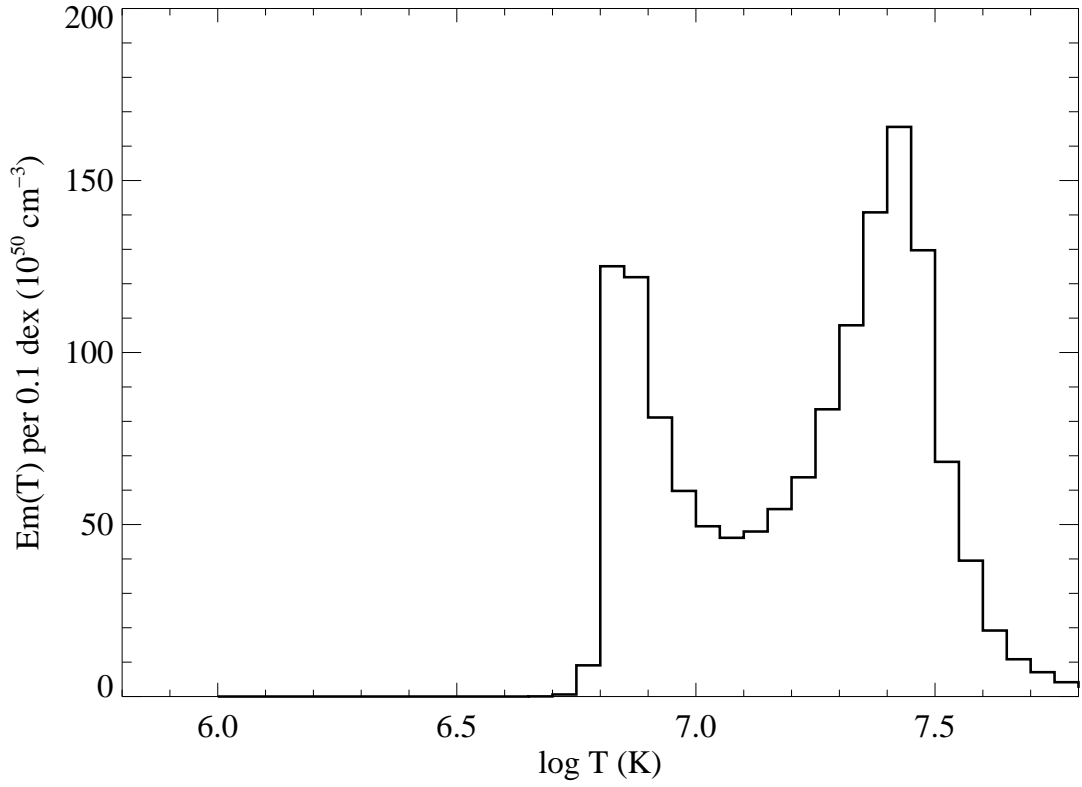}
\caption{
Emission-measure distribution produced by the microflaring model if
we assume $\alpha = 1.8$, compact loops, with a semi-length of
$2.6\times 10^9$~cm, and full surface coverage.
In this case, the equilibrium temperature is $T \approx 6\times 10^6$~K
and a dominant peak occurs near $T \approx 26\times 10^6$~K.}
\end{figure}

Our best-fit microflare model for HD~35850 uses 1100 loops,
distributed evenly over the star's surface, and accommodating
$2.64\times 10^7$ flares during a typical observation lasting
$5\times 10^5$~s.
The resulting EM distribution is shown in Figure~14.
The soft X-ray light curves from these simulations show the same 1--12~hour
variability seen in {\it ASCA} SIS and {\it EUVE} DS light curves.  
Because the power-law distribution of flares favors small (microflare) events,
the simulated light curves also show lower-amplitude, shorter-time scale
variability seen in higher-S/N X-ray observations of dMe stars but not
discernible for HD~35850.

The model EM distribution exhibits the characteristic
two-temperature structure with a broad intermediate minimum.
However, in this model the EM structure is not just a result
of radiative cooling as suggested by Gehrels \& Williams (1993).
It is a consequence of the balance between flare heating,
cooling by conduction, and cooling by radiation.
Given our simplistic model, it is noteworthy that the 
two-temperature structure (or rather the broad
minimum around 10~MK) observed on many active normal stars
can be explained, in part, by a hydrodynamic property of frequently-flaring
loops.

\section{The Active Corona of HD~35850}

We summarize the combined {\it EUVE} and {\it ASCA} results for
HD~35850 in Table~3.
The {\it ASCA} SIS spectral analysis suggests sub-solar abundances
and the {\it EUVE} line-to-continuum ratio
indicates approximately solar photospheric abundances.
Although $Z \approx 0.8$
is marginally consistent with the {\it EUVE} and {\it ASCA} spectra,
all the {\it ASCA} analyses yield systematically lower Fe abundance than
the {\it EUVE} Fe line-to-continuum ratio.

Assuming photospheric abundances, the corona is characterized by
emission measure in a warm component from 5--8~MK
and a hot component from 21--30~MK.
Like other rapidly rotating, Pleiades-age dwarfs,
HD~35850 appears to represent an activity
extremum for main-sequence solar-type stars.
HD~35850's inverse Rossby number is
$\tau_{\rm c}/P_{\rm rot} \approx 20.7/1.40 \approx 15$ 
and its X-ray surface flux is
$F_{\rm X} \approx 1.8 \times 10^{7}$~ergs~s$^{-1}$~cm$^{-2}$.
For AB~Dor (K0~V), 
$F_{\rm X} \approx 1.7 \times 10^{7}$~ergs~s$^{-1}$~cm$^{-2}$
(Hempelmann et al. 1995) and $\tau_{\rm c}/P_{\rm rot} > 120$.
For EK~Dra (G0~V),
$F_{\rm X} \approx 1.5 \times 10^{7}$~ergs~s$^{-1}$~cm$^{-2}$
and $\tau_{\rm c}/P_{\rm rot} \approx 15$.
AB Dor, HD~35850, and EK~Dra appear to have saturated or nearly 
saturated X-ray activity and,
despite having more rapid rotation and a deeper convection zone,
AB~Dor's X-ray surface flux is the same as HD~35850's.  

The {\it EUVE} DS light curve of HD~35850 shows about one moderate-amplitude
(40\% increase) flare per day.
The X-ray and EUV variability and the presence
of substantial emission measure above 20~MK suggests that some flare-like
mechanism must be heating the corona.
To test this hypothesis, we have
modeled HD~35850's light curves and ${\rm Em}(T)$ using the hydrodynamic
microflare point model of G\"udel (1997).
The simulations suggest a power-lax index of flare energies
$\alpha \approx 1.8$.

On the Sun, transient brightenings seen in {\it Yohkoh}
Soft X-ray Telescope images of solar active regions provide a direct
measure of the microflaring amplitude and power-law index.
Microflaring can provide, at most, 20\% of the heating rate
required to power the active-region corona (Shimizu 1995).
Ofman, Davila, \& Shimizu (1996) have shown that the transient brightenings seen
by {\it Yohkoh} may be a consequence of resonant absorption of
global-mode Alfv\'en waves
in coronal loops, excited by random footpoint motions of these loops.
A more recent analysis of {\it Yohkoh} data suggests that a
steeper distribution of smaller flares, the so-called nanoflares,
may also be present (Shimizu \& Tsuneta 1997).
Whether the microflares occur as a result of Alfv\'en waves
or magnetic reconnection, they cannot account for the radiative output of
the Solar corona.  For the rapidly rotating F and G dwarfs, however,
the microflare models are able to reproduce the observed luminosity,
EM distribution, and variability.

To better test microflaring on stars,
accurate coronal density and abundance measurements as a function
of temperature are needed to better constrain the emission-measure distribution.
Photospheric magnetic field strength and filling fractions can further
constrain loop models.
Also, higher S/N, higher cadence hard and soft X-ray light curves 
will provide a statistically robust estimate of the distribution of flares.
Planned observations of AB~Dor with the
{\it Advanced X-ray Astrophysics Facility}'s
High-Energy Transmission Grating Spectrometer will help resolve some of
the density and abundance issues discussed in this paper
and may lead to more quantitative tests of coronal heating models.

\acknowledgements

MG would like to thank Tom Ayres for providing some of the IDL routines to
perform optimal extraction and line fitting.  
MG made extensive use of
the NASA Astrophysics Data System abstract service,
the HEASARC database at NASA/GSFC, and
the Simbad database at the Centre de Donn\'ees astronomiques de Strasbourg.
The authors would like to thank an anonymous referee for many helpful
suggestions.
This research was supported under NASA grant NAG-2891 to the
University of Colorado.

%\documentstyle[apjpt4]{article}
%\pagestyle{empty}
%\begin{document}

\begin{planotable}{llcccccc}
\tablenum{1}
\tablewidth{0pc}
\tablecaption{HD 35850: 1995 October 23--30: EUVE SW Lines}
\tablehead{
\colhead{Dominant} &
\colhead{Blended} &
\colhead{$\log T$} &
\colhead{$\lambda_{\rm lab}$} &
\colhead{$\lambda_{\rm obs}$} &
\colhead{Detect} &
\colhead{$L_{\rm line}$} &
\colhead{S/N} \nl
\colhead{Ion} &
\colhead{Ion} &
\colhead{(K)} &
\colhead{(\AA)} &
\colhead{(\AA)} &
\colhead{Flag} &
\colhead{($10^{27}$~ergs~s$^{-1}$)} &
\colhead{} }
\startdata
\ion{Fe}{15} &                           & 6.3 & \phn73.47 & \nodata &   N &       $<$6.00 & \nodata \nl 
\ion{Fe}{13} &                           & 6.2 & \phn74.85 & \nodata &   N &       $<$4.43 & \nodata \nl 
\ion{Fe}{13} &             \ion{Fe}{13}  & 6.2 & \phn75.89 & \nodata &   N &       $<$4.47 & \nodata \nl 
\ion{Fe}{12} &                           & 6.2 & \phn79.49 & \nodata &   N &       $<$2.40 & \nodata \nl 
\ion{Fe}{12} &                           & 6.2 & \phn80.02 & \nodata &   N &       $<$2.13 & \nodata \nl 
\ion{Fe}{12} &                           & 6.2 & \phn80.51 & \nodata &   N &       $<$2.14 & \nodata \nl 
\ion{Fe}{11} &             \ion{Fe}{11}  & 6.1 & \phn86.76 & \nodata &   N &       $<$1.28 & \nodata \nl 
 \ion{Ne}{8} &                           & 5.8 & \phn88.08 & \nodata &   N &       $<$1.20 & \nodata \nl 
\ion{Fe}{11} &             \ion{Fe}{11}  & 6.1 & \phn89.18 & \nodata &   N &       $<$1.08 & \nodata \nl 
\ion{Fe}{19} &                           & 6.9 & \phn91.02 & \nodata &   N &       $<$1.04 & \nodata \nl 
\ion{Fe}{18} & \ion{Fe}{20}              & 6.8 & \phn93.92 &\phn93.94&   Y &      \phs8.87 &12.08\phn\nl 
\ion{Fe}{10} &                           & 6.1 & \phn95.37 & \nodata &   N &       $<$0.90 & \nodata \nl 
\ion{Fe}{10} &                           & 6.1 & \phn96.12 & \nodata &   N &       $<$0.91 & \nodata \nl 
\ion{Fe}{10} &                           & 6.1 & \phn97.12 & \nodata &   N &       $<$0.91 & \nodata \nl 
\ion{Fe}{21} &  \ion{Ne}{8}              & 7.0 & \phn97.88 &\phn98.13&   Y &      \phs0.98 &    2.03 \nl 
\ion{Fe}{22} &                           & 7.1 &    100.78 &  100.70 &   Y &      \phs0.93 &    2.27 \nl 
\ion{Fe}{19} &                           & 6.9 &    101.55 &  101.56 &   Y &      \phs1.96 &    4.09 \nl 
\ion{Fe}{21} &                           & 7.0 &    102.22 &  102.26 &   Y &      \phs2.24 &    4.97 \nl 
\ion{Fe}{18} &                           & 6.8 &    103.94 &  103.90 &   Y &      \phs2.21 &    4.60 \nl 
\ion{Fe}{19} &             \ion{Fe}{21}  & 6.9 &    108.37 &  108.35 &   Y &      \phs8.49 &13.17\phn\nl 
\ion{Fe}{19} &                           & 6.9 &    109.97 &  109.94 &   Y &      \phs1.36 &    3.24 \nl 
\ion{Fe}{19} &             \ion{Ni}{23}  & 6.9 &    111.70 &  111.63 &   Y &      \phs1.27 &    2.93 \nl 
\ion{Fe}{22} &                           & 7.1 &    114.41 &  114.42 &   Y &      \phs1.79 &    3.94 \nl 
  \ion{O}{6} &                           & 5.5 &    115.83 & \nodata &   N &       $<$0.83 & \nodata \nl 
\ion{Fe}{22} &             \ion{Fe}{21}  & 7.1 &    117.17 &  117.21 &   Y &      \phs3.83 &    6.48 \nl 
\ion{Ni}{25} &                           & 7.2 &    117.95 &  118.12 &   Y &      \phs1.14 &    2.62 \nl 
\ion{Fe}{20} &                           & 7.0 &    118.66 &  118.85 &   Y &      \phs2.99 &    5.68 \nl 
\ion{Fe}{19} &                           & 6.9 &    120.00 &  119.96 &   Y &      \phs2.07 &    4.01 \nl 
\ion{Fe}{20} &                           & 7.0 &    121.83 &  121.88 &   Y &      \phs4.87 &    8.37 \nl 
\ion{Fe}{21} &                           & 7.0 &    128.73 &  128.86 &   Y & \phs10.03\phn &11.41\phn\nl 
  \ion{O}{6} &                           & 5.5 &    129.87 & \nodata &   N &       $<$1.10 & \nodata \nl 
\ion{Fe}{23} &             \ion{Fe}{20}  & 7.1 &    132.85 &  132.94 &   Y & \phs24.61\phn &19.19\phn\nl 
\ion{Fe}{22} &                           & 7.1 &    135.78 &  135.89 &   Y &      \phs4.93 &    6.27 \nl 
\ion{Ca}{12} &                           & 6.3 &    141.03 & \nodata &   N &       $<$1.52 & \nodata \nl 
\ion{Fe}{21} &                           & 7.0 &    142.27 & \nodata &   N &       $<$1.63 & \nodata \nl 
\ion{Ni}{11} &                           & 6.2 &    148.37 & \nodata &   N &       $<$1.66 & \nodata \nl 
  \ion{O}{6} &                           & 5.5 &    150.09 & \nodata &   N &       $<$1.82 & \nodata \nl 
\ion{Ni}{12} &                           & 6.2 &    152.15 & \nodata &   N &       $<$1.95 & \nodata \nl 
\ion{Ni}{12} &                           & 6.2 &    154.18 & \nodata &   N &       $<$2.06 & \nodata \nl 
\ion{Ni}{13} &             \ion{Ni}{10}  & 6.3 &    157.73 & \nodata &   N &       $<$2.22 & \nodata \nl 
\ion{Ni}{12} &             \ion{Ni}{13}  & 6.3 &    160.50 & \nodata &   N &       $<$2.68 & \nodata \nl 
\ion{Ar}{13} &             \ion{Ca}{13}  & 6.4 &    161.61 & \nodata &   N &       $<$2.64 & \nodata \nl 
\ion{Ni}{13} &             \ion{Ar}{13}  & 6.3 &    164.15 & \nodata &   N &       $<$3.04 & \nodata \nl 
\ion{Ar}{10} &                           & 6.1 &    165.51 & \nodata &   N &       $<$2.81 & \nodata \nl 
\ion{Ni}{14} &                           & 6.3 &    169.68 & \nodata &   N &       $<$3.14 & \nodata \nl 
\enddata
\end{planotable}

%\end{document}

%\documentstyle[apjpt4]{article}
%\pagestyle{empty}
%\begin{document}

\begin{planotable}{llcccccc}
\tablenum{2}
\tablewidth{0pc}
\tablecaption{HD 35850: 1995 October 23--30: EUVE MW Lines}
\tablehead{
\colhead{Dominant} &
\colhead{Blended} &
\colhead{$\log T$} &
\colhead{$\lambda_{\rm lab}$} &
\colhead{$\lambda_{\rm obs}$} &
\colhead{Detect} &
\colhead{$L_{\rm line}$} &
\colhead{S/N} \nl
\colhead{Ion} &
\colhead{Ion} &
\colhead{(K)} &
\colhead{(\AA)} &
\colhead{(\AA)} &
\colhead{Flag} &
\colhead{($10^{27}$~ergs~s$^{-1}$)} &
\colhead{} }
\startdata
 \ion{Fe}{9} &                           & 5.8 &    171.07 &  170.78 &   Y &      \phs4.78 &    2.25 \nl 
  \ion{O}{6} &                           & 5.5 &    173.09 & \nodata &   N &       $<$3.45 & \nodata \nl 
  \ion{O}{6} &                           & 5.5 &    183.95 & \nodata &   N &       $<$3.43 & \nodata \nl 
\ion{Fe}{24} &             \ion{Ca}{17}  & 7.2 &    192.04 &  191.62 &   Y & \phs14.47\phn &    7.28 \nl 
\ion{Ar}{14} &             \ion{Ca}{14}  & 6.5 &    194.41 &  193.31 &   Y &      \phs5.86 &    3.08 \nl 
\ion{Fe}{14} &                           & 6.3 &    211.33 & \nodata &   N &       $<$4.44 & \nodata \nl 
\ion{Fe}{14} &                           & 6.3 &    219.13 & \nodata &   N &       $<$4.07 & \nodata \nl 
\ion{Fe}{15} &                           & 6.3 &    243.80 & \nodata &   N &       $<$4.67 & \nodata \nl 
\ion{Fe}{16} &                           & 6.4 &    251.04 & \nodata &   N &       $<$5.19 & \nodata \nl 
\ion{Fe}{24} &             \ion{Fe}{17}  & 7.2 &    255.10 &  255.10 &   Y &      \phs7.68 &    3.15 \nl 
 \ion{He}{2} &              \ion{S}{13}  & 6.9 &    256.30 &  255.64 &   N &       $<$7.68 & \nodata \nl 
\ion{Fe}{16} &             \ion{Fe}{23}  & 6.4 &    262.97 &  262.76 &   Y &      \phs5.42 &    2.63 \nl 
\ion{Fe}{14} &                           & 6.3 &    264.78 & \nodata &   N &       $<$4.37 & \nodata \nl 
\ion{Fe}{15} &                           & 6.3 &    284.15 &  284.38 &   Y & \phs13.27\phn &    6.04 \nl 
\ion{Ni}{18} &                           & 6.5 &    292.00 & \nodata &   N &       $<$4.91 & \nodata \nl 
 \ion{He}{2} &             \ion{Si}{11}  & 6.6 &    303.91 &  303.87 &   N & \phs77.68\phn &16.50\phn\nl 
\ion{Fe}{16} &                           & 6.4 &    335.41 &  334.94 &   Y & \phs41.54\phn &13.86\phn\nl 
\ion{Fe}{16} &                           & 6.4 &    360.80 &  359.68 &   Y & \phs21.18\phn &    7.96 \nl

\enddata
\end{planotable}

%\end{document}

%\documentstyle[apjpt4]{article}
%\pagestyle{empty}
%\begin{document}

\begin{planotable}{lcl}
\tablenum{3}
\tablewidth{0pc}
\tablecaption{HD 35850: Derived and Published Parameters}
\tablehead{
\colhead{Parameter} &
\colhead{Value} &
\colhead{Reference} }
\startdata
MK			& F8--9~V							& Cutispoto et al. 1995 	\nl
$d$			& $26.84\pm0.62$~pc					& Perryman et al. 1997		\nl
$t_{\star}$		& $10^8$~yr							& Tagliaferri et al. 1994	\nl
$V$			& $6.300\pm0.006$						& Perryman et al. 1997		\nl
$B-V$			& $0.553\pm0.007$						& Perryman et al. 1997		\nl
$M$			& $1.15 M_{\odot}$					& Kim \& Demarque 1996		\nl
$\tau_{\rm c}$	& 20.7~d							& Kim \& Demarque 1996		\nl
$L_{\rm bol}$	& $6.64\times 10^{33}$~ergs~s$^{-1}$		& Blackwell \& Lynas-Gray 1994\nl
$R_{\star}$		& $1.18 R_{\odot}$					& Blackwell \& Lynas-Gray 1994\nl
$v\sin i$		& $50$~km~s$^{-1}$					& Tagliaferri et al. 1994	\nl
$P_{\rm rot}$	& $1.40\pm0.20$~d						& \nodata				\nl
${\rm Ro}^{-1}$	& $14.8\pm2.4$							& \nodata				\nl
$L_{\rm X}$		& $1.54\times 10^{30}$~ergs~s$^{-1}$		& \nodata				\nl
$R_{\rm X}$		& $2.3\times 10^{-4}$					& \nodata				\nl
$F_{\rm X}$		& $1.8\times 10^{7}$~ergs~cm$^{-2}$~s$^{-1}$	& \nodata				\nl
$N_{\rm H}$		& $1.7^{+0.3}_{-1.1}\times 10^{18}$~cm$^{-2}$		& \nodata				\nl
$Z$			& $1.15^{+0.75}_{-0.35}$						& \nodata				\nl
$T_1$			& $6.3^{+1.6}_{-2.3}$~MK				& \nodata				\nl
${\rm Em}_1$	& $2.8\times 10^{52}$~cm$^{-3}$			& \nodata				\nl
$T_2$			& $25.1^{+6.5}_{-5.2}$~MK				& \nodata				\nl
${\rm Em}_2$	& $4.0\times 10^{52}$~cm$^{-3}$			& \nodata				\nl
\enddata
\end{planotable}
%\end{document}


\begin{references}

\reference {} Anders, E. \& Grevesse, N. 1989, Geochimica et Cosmochimica Acta, 53, 197
\reference {} Arnaud, K.A. 1996, {\it Astronomical Data Analysis
Software and Systems V}, eds. G.H. Jacoby \& J. Barnes, ASP Conference
Series, 101, 17
\reference {} Baliunas, S.L., et al. 1995, ApJ, 438, 269
\reference {} Baliunas, S.L., Nesme-Ribes, E., Sokoloff, D., \& Soon,
W.H. 1996, 460, 848
\reference {} Benz, A.O., \& G\"udel, M. 1994, \aap, 285, 621
\reference {} Blackwell, D.E., \& Lynas-Gray, A.E. 1994, \aap, 282, 899
\reference {} Brickhouse, N.S., Raymond, J.C., \& Smith, B.W. 1995,
\apjs, 97, 551
\reference {} Brickhouse, N.S., Dupree, A.K., Edgar, R.J., Drake,
S.A., White, N.E., Liedahl, D.A., Singh, K.P. 1997, AAS, 191, 2513
\reference {} Cutispoto, G., Tagliaferri, G., Giommi, P., Gouiffes,
C., Pallavicini, R., Pasquini, L., \& Rodon\'o, M. 1991, \aaps, 87, 233
\reference {} Donati, J.-F., \& Collier Cameron, A. 1997, \mnras, 291, 1
\reference {} Feigelson. E.D., \& Babu, G.J. 1992, {\it Statistical
Challenges in Modern Astronomy}, (Springer-Verlag: New York)
\reference {} Field, G.B. 1965, \apj, 142, 531
\reference {} Gehrels, N., \& Williams, E.D. 1993, \apj, 418, L25
\reference {} Gray, D.F., \& Baliunas, S.L. 1997, ApJ, 475, 303
\reference {} G\"udel, M. Guinan, E.F., Mewe, R., Kaastra, J., \&
Skinner, S.L. 1997, ApJ, 483, 947
\reference {} G\"udel, M. 1997, \apj, 480, L121
\reference {} Haisch, B., Bowyer, S., \& Malina, R. 1993, JBIS, 46, 331
\reference {} Haisch, B., \& Schmitt, J.H.M.M. 1996, \pasp, 108, 113
\reference {} Halpern, J., \& Marshall, H. 1996, \apj, 464, 760
\reference {} Hempelmann, A., Schmitt, J.H.M.M., Schultz, M., R\"udiger, G., 
\& St\c{e}pie\`{n}, K. 1995, \aap, 294, 515
\reference {} Kaastra, J.S. 1992, {\it An X-ray spectral code for
optically thin plasmas}, internal SRON-Leiden Report, updated version 2.0
\reference {} Kim, Y.-C., \& Demarque, P. 1996, ApJ, 457, 340
\reference {} Kopp, R.A., \& Polletto, G. 1993, \apj, 418, 496
\reference {} K\"urster, M., Schmitt, J.H.M.M., Cutispoto, G., \&
Dennerl, K. 1997, \aap, 320, 831
\reference {} Lemen, J.R., Mewe, R., Schrijver, C., \& Fludra,
A. 1989, \apj, 341, 474
\reference {} Lenz, D.D., \& Ayres, T.R. 1992, \pasp, 104, 1104
\reference {} Mathioudakis, M., \& Mullan, D.J. 1998, \aap, in press
\reference {} Mewe, R., Kaastra, J., White, S.M., Pallavicini,
R. 1996, \aap, 315, 170
\reference {} Mewe, R., Lemen, J.R., Peres, G., Schrijver, C., Serio,
S., \aap, 152, 229
\reference {} Monsigniori Fossi, B., \& Landini, M. 1994, \solphys,
152, 81
\reference {} Noyes, R., Hartmann, L., Baliunas, S., Duncan, D., \& Vaughn, A. 1984, ApJ, 279, 763
\reference {} Ofman, L., Davila, J.M., \& Shimizu, T. 1996, ApJ, L39
\reference {} Oreshina, A.V. \& Somov, B.V. 1997, \aap, 320, L53
\reference {} Parker, E.N. 1953, \apj, 117, 431
\reference {} Perryman, M.A.C., et al. 1997, \aap, 323, L49
\reference {} Poletto, G., Pallavicini, R., \& Kopp, R.A. 1988, A\&A, 201, 93
\reference {} Rutten, R.G.M., Schrijver, C.J., Lemmens, A.F.P., \&
Zwaan, C. 1991, \aap, 252, 203
\reference {} Scargle, J.D. 1989, \apj, 343, 874
\reference {} Shimizu, T. 1995, PASJ, 47, 251
\reference {} Shimizu, T., \& Tsuneta, S. 1997, ApJ, 486, 1045
\apj, 296, 46
\reference {} Tagliaferri, G., Cutispoto, G., Pallavicini, R.,
Randich, S., \& Pasquini, L. 1994, \aap, 285, 272
\reference {} Tagliaferri, G., Covino, S., Fleming, T.A., Gagn\'e,
M. , Pallavicini, R., Haardt, F., \& Uchida, Y. 1997, \aap, 321, 850 
\reference {} Wood, B.E., Harper, G.M., Linsky, J.L., \& Dempsey,
R.C. 1996, ApJ, 458, 761

\end{references}
\end{document}